\documentclass[review]{elsarticle}

\usepackage{lineno,hyperref}
\usepackage{amssymb}
\usepackage{amsmath}
\usepackage{graphicx}
\usepackage{epstopdf, epsfig}
\usepackage{hyperref}
\usepackage{cleveref}
\usepackage{multirow}
\usepackage{color}

\let\sss= \scriptscriptstyle
\let\ss= \scriptstyle
\modulolinenumbers[5] 

\journal{ }

\begin{document}

\begin{frontmatter}

\title{NSFnets (Navier-Stokes Flow nets): Physics-informed neural networks for the incompressible Navier-Stokes equations}


\author[mymainaddress,mysecondaryaddress]{Xiaowei Jin}
\author[brown]{Shengze Cai}

\author[mymainaddress,mysecondaryaddress]{Hui Li\corref{mycorrespondingauthor}}
\ead{lihui@hit.edu.cn}
\author[brown]{George Em Karniadakis\corref{mycorrespondingauthor}}
\cortext[mycorrespondingauthor]{Corresponding authors}
\ead{george\_karniadakis@brown.edu}

\address[mymainaddress]{Key Lab of Smart Prevention and Mitigation of Civil Engineering Disasters of the Ministry of Industry and Information Technology, Harbin Institute of Technology, Harbin 150090, China}
\address[mysecondaryaddress]{Key Lab of Structures Dynamic Behavior and Control of the Ministry of Education, Harbin Institute of Technology, Harbin 150090, China}
\address[brown]{Division of Applied Mathematics, Brown University, Providence, RI 02912, USA}

\begin{abstract}
We employ physics-informed neural networks (PINNs) to simulate the incompressible flows ranging from laminar to turbulent flows. We perform PINN simulations by considering two different formulations of the Navier-Stokes equations: the velocity-pressure (VP) formulation and the vorticity-velocity (VV) formulation. We refer to these specific PINNs for the Navier-Stokes flow nets as NSFnets. Analytical solutions and direct numerical simulation (DNS) databases provide proper initial and boundary conditions for the NSFnet simulations. The spatial and temporal coordinates are the inputs of the NSFnets, while the instantaneous velocity and pressure fields are the outputs for the VP-NSFnet, and the instantaneous velocity and vorticity fields are the outputs for the VV-NSFnet. These two different forms of the  Navier-Stokes equations together with the initial and boundary conditions are embedded into the loss function of the PINNs. No data is provided for the pressure to the VP-NSFnet, which is a hidden state and is obtained via the incompressibility constraint without splitting the equations. We obtain good accuracy of the NSFnet simulation results upon convergence of the loss function, verifying that NSFnets can effectively simulate complex  incompressible flows using either the VP or the VV formulations. For the  laminar flow solutions we show that the VV formulation is more accurate than the VP formulation. For the turbulent channel flow we show that NSFnets can sustain turbulence at  $ \rm Re_\tau \sim  1,000$ but due to expensive training we only consider part of the channel domain and enforce velocity boundary conditions on the boundaries provided by the DNS data base. We also perform a systematic study on the weights used in the loss function for the data/physics components and investigate a new way of computing the weights dynamically to accelerate training and enhance accuracy. 
Our results suggest that the accuracy of NSFnets, for both laminar and turbulent flows, can be improved with proper tuning of weights (manual or dynamic) in the loss function. 
\end{abstract}


\end{frontmatter}


\section{Introduction}

In the last five years there have been several efforts to integrate neural networks (NNs) in the solution of the incompressible Navier-Stokes equations following different approaches. For turbulent flows, the most common approach is to derive data-driven turbulence closure models. For example, \citet{ling2016reynolds} proposed a data-driven Reynolds-averaged Navier–Stokes (RANS) turbulence closure model by embedding Galilean invariance into deep neural networks and demostrated better accuracy for  predicting the Reynolds stresses. Similarly, \citet{wang2017physics} used random forest regression to predict the discrepancies of the baseline RANS-predicted Reynolds stresses compared to those from the DNS data, hence predicting the  Reynolds stresses with high accuracy. \citet{jiang2020novel} developed a novel RANS stress closure with machine-learning-assisted parameterization and nonlocal effects, aiming at reducing both structural and parameteric inaccuracies and achieving a more appropriate description for Reynolds stress anistropy. For large-eddy simulation (LES) of isotropic turbulence, \citet{zhou2019subgrid} developed a data-driven subgrid scale model by using NNs with only one hidden layer. In addition, some reduced order models (ROMs) or fast prediction models in fluid mechanics have also been investigated. For example, convolutional neural networks (CNNs) were used to construct the prediction model of cylinder wake  in \citep{jin2018prediction}, and a temporal CNN was used to establish a data-driven model for predicting the coefficients of proper orthogonal decomposition (POD) modes of cylinder wake in \citep{wu2020data}. The bidirectional recurrent neural networks were employed to predict the POD coefficients of cylinder wake based on a few velocity measurements \citep{jin2020time}, obtaining more accurate results than the extended POD approach \citep{hosseini2015sensor, discetti2018estimation}. Moreover, deep learning techniques were also applied to particle image velocimetry (PIV) for analyzing laboratory data of turbulent boundary layer \citep{cai2019dense}.
Comprehensive summaries of progress in fluid mechanics due to the introduction of  various machine learning techniques can be found in \citep{duraisamy2019turbulence, brunton2020machine}. 

We have followed a different path by exploiting the universal approximation property of NNs, which together with automatic differentiation enables us to develop Navier-Stokes ``solvers" that do not require mesh generation. They are easy to implement, and can be particularly effective for multiphysics and inverse fluid mechanics problems. In particular, \citet{raissi2017physicsA,raissi2017physicsB, raissi2019physics} first introduced the concept of physics-informed neural networks (PINNs) to solve forward and inverse problems involving several different types of PDEs. This approach has also been used to simulate vortex induced vibrations in \cite{raissi2019deep} and also to tackle ill-posed inverse fluid mechanics problems, a framework called ``hidden fluid mechanics" presented in \cite{raissi2020hidden}. The flows considered in the aforementioned works are laminar flows at relatively low Reynolds numbers, described by the incompressible Navier-Stokes equations in velocity-pressure (VP) form. A fundamental question is if PINNs can simulate turbulence directly, similarly to direct numerical simulation (DNS) using high-order discretization \cite{kim1987turbulence,karniadakis2013spectral}. Another important question is if there is another formulation of the Navier-Stokes equations, e.g., in vorticity-velocity (VV) form, that may achieve higher accuracy or may be amenable to a more efficient training.  

In the current study, we address the aforementioned two questions systematically by using analytical solutions for two-dimensional and three-dimensional flows and also by comparing with DNS of turbulent channel flow available at \citep{perlman2007data,li2008public,graham2016web}. In particular, we perform PINN simulations by considering two forms of the governing Navier-Stokes equations: the VP form and the VV form, and we refer to these PINNs for the Navier-Stokes flow nets as NSFnets. For the VP-NSFnet, the inputs are the spatial and temporal coordinates while the outputs are the instantaneous velocity and pressure fields. For the VV-NSFnet, the inputs again are the spatial and temporal coordinates while the outputs are the instantaneous velocity and vorticity fields. We use automatic differentiation (AD) \citep{baydin2018automatic} to deal with the differential operators in the Navier-Stokes equations, which leads to very high computational efficiency compared to numerical differentiation. However, it does not require grids, and avoids the classical artificial dispersion and diffusion errors. Furthermore, with AD we differentiate the NN rather than the data directly and hence we can deal with noisy inputs or solutions with limited regularity. There are also distinct advantages in employing both the VP and the VV formulations using PINNs. For example, to infer the pressure equation we do not use an additional Poisson pressure equation as is usually done with the traditional splitting  methods \citep{karniadakis1991high} and no data is required for the pressure as boundary or initial conditions for VP-NSFnet; the pressure is a hidden state and is obtained via the incompressibility constraint. Similarly, in the VV-NSFnet, it is easy to incorporate vorticity boundary conditions, which come in the form of constraints, directly into the loss function.

We simulate several laminar flows, including two-dimensional steady Kovasznay flow, two-dimensional unsteady cylinder wake and three-dimensional unsteady Beltrami flow, using these two types of NSFnets. We perform a systematic study using dynamic weights in the loss function for the various components following the work of \citep{wang2020understanding} to accelerate training and enhance accuracy. We also report the first results on directly simulating turbulence using PINNs. To this end, we consider 
turbulent channel flow at $ \rm Re_\tau \sim $ 1,000 using primarily VP-NSFnet as the available data bases are derived based on
VP type formulations. We perform NSFnet simulations by considering different subdomains with different size at various locations in the channel and for different time intervals. In addition, we investigate the influence of weights in the loss function on the accuracy of VP-NSFnet. 

The paper is organized as follows. We first introduce the NSFnets in \cref{sec:Methodology}, and present the problem set up and NSFnet simulation results for laminar flows in \cref{sec:analyticalExamples}. We then present VP-NSFnet results  for turbulent channel flow in \cref{sec:turbulentExamples}. 
We summarize our findings in \cref{sec:Conclusion}.

\section{Solution Methodology}\label{sec:Methodology}
\begin{figure}
	\centerline{\includegraphics [width=12cm] {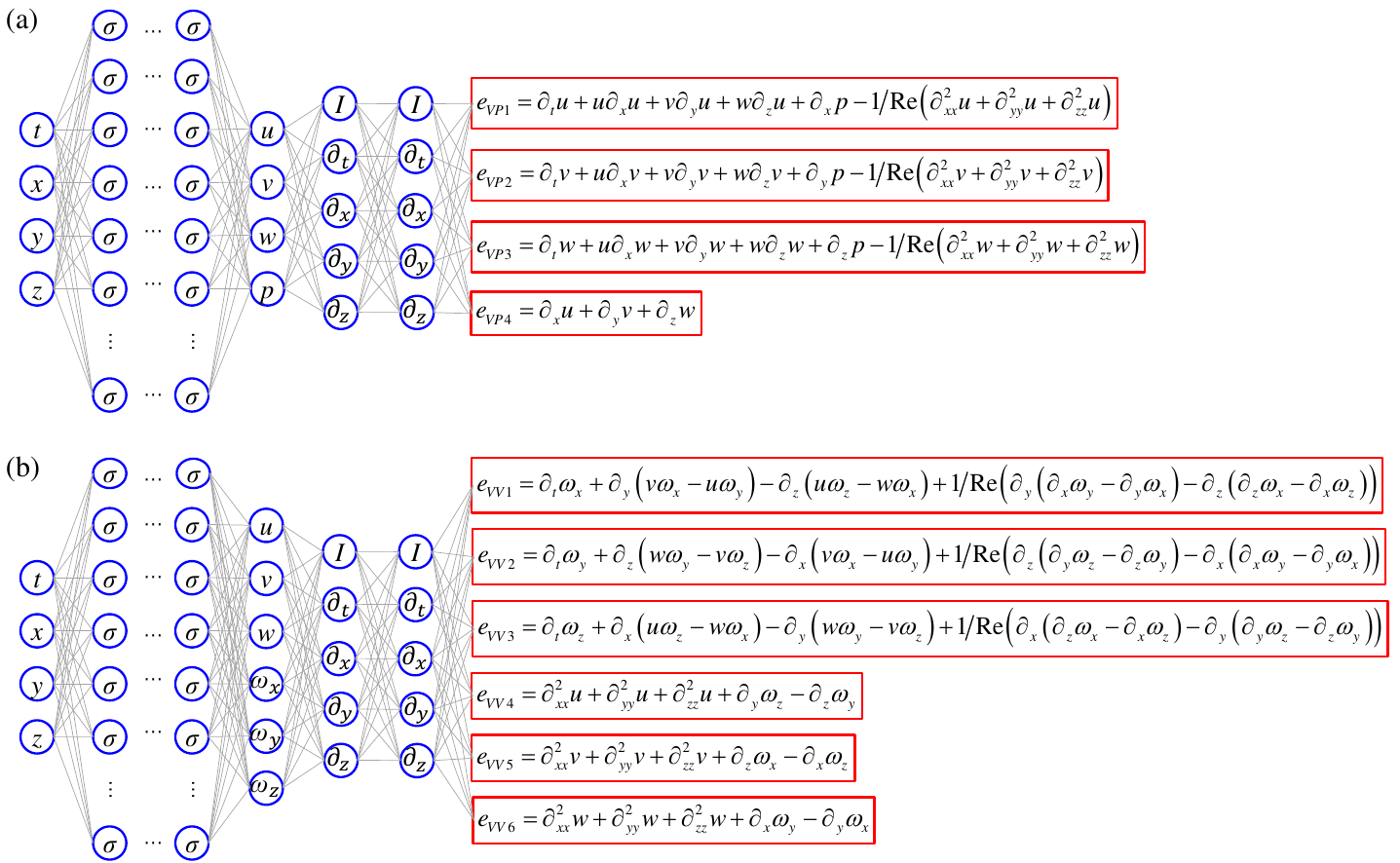}}
	\caption{A schematic of NSFnets: (a) the velocity-pressure (VP) form; (b) the vorticity-velocity (VV) form. The left part of the NN is an uninformed network, while the right part implements the VP and VV formulation using AD. We only show the operators in the right part as the NNs induced by AD of the VP and VV differential operators are too complicated and cannot be visualized even with specialized methods such as ``TensorBoard" in TensoFlow. }
	\label{fig:neutral_networks}
\end{figure}

We introduce two formulations of the unsteady incompressible three-dimen\-sional Navier-Stokes equations: the velocity-pressure (VP) form and the vorticity-velocity (VV) form, as well as their corresponding physics-informed neural networks (PINNs), shown in
Fig. \ref{fig:neutral_networks}. 

The VP form of the incompressible Navier-Stokes equations is:
\begin{subequations}\label{e:NS_vel_pres}
	\begin{align}
		\frac{{\partial \mathbf{u}}}{{\partial t}} + (\mathbf{u}\cdot\nabla)\mathbf{u} &= -\nabla p + \frac{1}{\text{Re}}\nabla^{2}\mathbf{u} \quad \quad  \text{in} ~\Omega ,\\
		\nabla \cdot \mathbf{u}  &= 0  \quad \quad  \text{in} ~\Omega , \\
		\mathbf{u}  &=  \mathbf{u}_{\Gamma} \quad \quad  \text{on}~ \Gamma_{D},  \\
		\frac{{\partial \mathbf{u}}}{{\partial n}} &=0 \quad\quad   \text{on} ~ \Gamma_{N},
	\end{align}
\end{subequations}
where $ t $ is the non-dimensional time, $ \mathbf{u}(\mathbf{x},t) = [u,v,w]^{T}$ is the non-dimensional velocity vector, $ p $ is the non-dimensional pressure, and $ \text{Re} = U_{ref}D_{ref}/\nu $ is the Reynolds number defined by a characteristic length $ D_{ref} $, reference velocity $ U_{ref} $ and kinematic viscosity $ \nu $. The initial and boundary conditions are required in order to solve Eq. (\ref{e:NS_vel_pres}). Here, $\Gamma_{D}$ and $\Gamma_{N}$ denote the Dirichlet and Neumann boundaries, respectively. In this study, instead of using conventional computational fluid dynamics (CFD) methods, we investigate the possibility of using neural networks (NNs) for solving the Navier-Stokes equations. In other words, the solutions of Navier-Stokes equations are approximated by a deep neural network, which takes spatial and temporal coordinates as inputs and predicts the corresponding velocity and pressure fields, i.e., $(t,x,y,z)\mapsto (u,v,w,p)$. 
A schematic illustration of the PINNs for solving Eq. (\ref{e:NS_vel_pres}) is shown in Fig. \ref{fig:neutral_networks}a, which consists of a fully-connected network and the residual networks. Here, the nonlinear activation function $\sigma$ is the hyper tangent function \emph{tanh}. For the VP form, the residuals include the errors of the momentum equations and the divergence-free constraint. In order to compute the residuals of the Navier-Stokes equations $e_{\sss VP \ss 1}$ to $e_{\sss VP \ss 4}$, the partial differential operators are computed by using automatic differentiation (AD), which can be directly formulated in the deep learning framework, e.g., using ``\texttt{tf.gradients()}" in TensorFlow.

The loss function for training the parameters of VP-NSFnet to obtain the solutions of Eq. (\ref{e:NS_vel_pres}) is defined as follows:
\begin{subequations}\label{e:loss_vp}
	\begin{align}
		L &= L_e + \alpha L_b + \beta L_i \label{e:loss_vp_L1},   \\
		L_e &= \frac{1}{N_e}\sum_{i=1}^{4} \sum_{n=1}^{N_{e}} |e_{\sss VP\ss i}^{n} |^{2} \label{e:loss_vp_Le},   \\
		L_b & = \frac{1}{N_b}\sum_{n=1}^{N_{b}}  | \mathbf{u}^{n}-\mathbf{u}_{b}^{n}|^{2} \label{e:loss_vp_Lb},    \\
		L_i & = \frac{1}{N_i}\sum_{n=1}^{N_{i}}  | \mathbf{u}^{n}-\mathbf{u}_{i}^{n}|^{2}  \label{e:loss_vp_Ld},    
	\end{align}
\end{subequations}
where $ L_e $, $ L_b $ and $ L_i $ represent loss function components corresponding to the residual of the Navier-Stokes equations, the boundary conditions, and the initial conditions, respectively; $ N_b $, $ N_i $ and $ N_e $ denote the number of training data for different terms; $\mathbf{u}^{n}_{b} = [u^{n}_{b}, v^{n}_{b}, w^{n}_{b} ]^{T}$ and $\mathbf{u}^{n}_{i} = [u_{n}^{i}, v_{n}^{i}, w_{n}^{i}]^{T}$ are the given velocities for the $n$ th data point on the boundaries and at the initial time, respectively; $e_{\sss VP \ss i}^{n}$ represents the residual of the $i$ th equation at the $n$ th data point. 
The weighting coefficients $ \alpha $ and $ \beta $ are used to balance different terms of the loss function and accelerate convergence in the training process. We consider the initial and boundary conditions as supervised data-driven parts, and the residual of the Navier-Stokes equations as the unsupervised physics-informed part in the loss function. We note that no data is provided for the pressure as boundary or initial conditions, which means that $p$ is a hidden state and is obtained via the incompressibility constraint without splitting the Navier-Stokes equations as done in traditional CFD methods \citep{karniadakis2013spectral}. An adaptive optimization algorithm, \emph{Adam} \citep{kingma2014adam}, is used to minimize the loss function in (\ref{e:loss_vp}). 
The parameters of the neural networks are randomly initialized using the Xavier scheme \citep{glorot2010understanding}. 
The solutions are obtained when the training of the NSFnet converges, i.e., the total loss function reaches some very small value.

We also propose NSFnets for the VV formulation of the Navier-Stokes equations, which is an alternative to the VP form in simulating incompressible flows; the equivalence of VP and VV formulations was proved in \citep{trujillo1999penalty,meitz2000compact}. The rotational form of the VV formation of the Navier-Stokes equations is:
\begin{subequations}\label{e:NS_vort_vel}
	\begin{align}
		\frac{{\partial \omega }}{{\partial t}} + \nabla \times (\mathbf{\omega}\times\mathbf{u}) &= - \frac{1}{\text{Re}}\nabla \times \nabla \times  \mathbf{\omega} \quad\quad   \text{in} ~\Omega, \label{e:NS_vv_a}\\
		\nabla^{2} \mathbf{u}  &= -\nabla \times \mathbf{\omega}   \quad\quad   \text{in} ~\Omega , \label{e:NS_vv_b}\\
		\mathbf{\omega}  &= \nabla \times \mathbf{u}  \quad\quad   \text{on}~ \Gamma ,  \label{e:NS_vv_c} \\
		\mathbf{u}  &=  \mathbf{u}_{\Gamma} \quad\quad   \text{on}~ \Gamma_{D} ,  \label{e:NS_vv_d} \\
		\frac{{\partial \mathbf{u}}}{{\partial n}}   &=0 \quad\quad  \text{on} ~ \Gamma_{N},    \label{e:NS_vv_e}  \\
		\nabla \cdot \mathbf{u} &= 0 \quad\quad  \text{at one point on} ~ \Gamma,  \label{e:NS_vv_f}  \\
		\mathbf{\omega}  &= \nabla \times \mathbf{u}  \quad  \text{at} ~ t=0 \quad \quad  \text{in} ~ \Omega,  \label{e:NS_vv_g} 
	\end{align}
\end{subequations}
where $\mathbf{\omega}=[\omega_{x}, \omega_{y}, \omega_{z}]^{T}$ is the vorticity with three components. The boundary conditions are defined by Eqs. (\ref{e:NS_vv_c}) to (\ref{e:NS_vv_f}) and the initial condition is constrained by Eq. (\ref{e:NS_vv_g}). 
Similarly, we assume that the solutions of the VV form (\ref{e:NS_vort_vel}) are approximated by a neural network, whose function can be written as $(t,x,y,z)\mapsto (u,v,w,\omega_{x}, \omega_{y}, \omega_{z})$.
The architecture of the VV-NSFnet for solving Eq. (\ref{e:NS_vort_vel}) is shown in Fig. \ref{fig:neutral_networks}b, where $e_{\sss VV \ss 1}$ to $e_{\sss VV \ss 6}$ represent the residuals of the VV formulation of the Navier-Stokes Eqs. (\ref{e:NS_vv_a}) and (\ref{e:NS_vv_b}). 
The corresponding loss function of the VV-NSFnet is defined as follows:
\begin{subequations}\label{e:loss_vv}
	\begin{align}
		L &= L_e + \alpha L_b + \beta L_i      \\
		L_e &= \frac{1}{N_e}\sum_{i=1}^{6} \sum_{n=1}^{N_{e}} |e_{\sss VV \ss i}^{n} |^{2}   \\
		L_b & = \frac{1}{N_b}\sum_{n=1}^{N_{b}}  \left( | \mathbf{u}^{n}-\mathbf{u}^{n}_{b}|^{2} + |\mathbf{\omega}^{n} - \nabla \times \mathbf{u}^{n}_{b}  |^{2} + |\nabla\cdot \mathbf{u}^{n}_{b} |^{2} \right)   \label{e:loss_vv_c}  \\
		L_i & = \frac{1}{N_i}\sum_{n=1}^{N_{i}}  \left( | \mathbf{u}^{n}-\mathbf{u}^{n}_{i}|^{2} + | \mathbf{\omega}^{n} - \nabla \times \mathbf{u}^{n}_{i}   |^{2} \right) , \label{e:loss_vv_d}
	\end{align}
\end{subequations}
where $\mathbf{\omega}^n=[\omega_{x}^{n}, \omega_{y}^{n}, \omega_{z}^{n}]^{T}$ denotes the vorticity  for the $n$ th data point by NSFnet; $e_{\sss VV \ss i}^{n}$ represents the residual of the $i$ th equation at the $n$ th data point. Note that only boundary and initial values of velocity are provided in the loss function. For the vorticity term, the boundary and initial conditions are embedded in the losses (\ref{e:loss_vv_c}) and (\ref{e:loss_vv_d}) as constraints. The parameters of the neural network are also learned by using the \emph{Adam} optimizer.


Note that the weighting coefficients in the loss functions (\ref{e:loss_vp}) and (\ref{e:loss_vv}) play a very important role in the training process. However, choosing appropriate weights for NSFnets is generally very tedious. On the one hand, the optimal values of $\alpha$ and $\beta$ are problem-dependent and we cannot fix them for different flows. On the other hand, tuning the weights arbitrarily requires a trial and error procedure which is quite tedious and time-consuming. To tackle this problem, we apply the strategy of dynamic weights \citep{wang2020understanding} for choosing $\alpha$ and $\beta$ in NSFnet simulations. The idea of dynamic weights is to adaptively update the coefficients by utilizing the back-propagated gradient statistics during network training. 
For a general gradient decent algorithm, the iterative formulation of the parameters of NSFnets can be expressed as:
\begin{equation}\label{e:theta_iterations}
    \theta^{(k+1)} = \theta^{(k)} - \eta  \nabla_{\theta}L_e - \eta \alpha \nabla_{\theta}L_b -  \eta \beta \nabla_{\theta}L_i, 
\end{equation}
where $\theta$ denotes the parameters of the neural network, namely the weights of all the fully-connected layers, $k$ is the iteration step, and $\eta$ is the learning rate. In order to balance the contributions of different terms in Eq. (\ref{e:theta_iterations}), \citet{wang2020understanding} proposed to use a dynamic weight strategy during network training. 
At each training step, e.g., $(k+1)$ th iteration, the estimates of $\alpha$ and $\beta$ can be computed by:
\begin{equation}\label{e:dyW_1}
    \hat{\alpha}^{(k+1)} = \frac{\max_{\theta}\{|\nabla_{\theta}L_e |\}}{ \overline{|\nabla_{\theta} {\alpha}^{(k)} L_b|} },  \quad
    \hat{\beta}^{(k+1)} = \frac{\max_{\theta}\{|\nabla_{\theta}L_e |\} }{ \overline{|\nabla_{\theta} {\beta}^{(k)} L_i|} }, 
\end{equation}
where $\max_{\theta}\{|\nabla_{\theta}L_e |\}$ is the maximum value attained by $ |\nabla_{\theta}L_e|$; 
$\overline{|\nabla_{\theta} {\alpha}^{(k)} L_b|}$ and $\overline{|\nabla_{\theta} {\beta}^{(k)} L_i|}$ denote the means of $ |\nabla_{\theta} {\alpha}^{(k)} L_b| $ and $ |\nabla_{\theta} {\beta}^{(k)} L_i| $, respectively. As an alternative, we also propose the following way to estimate $\alpha$ and $\beta$:
\begin{equation}\label{e:dyW_mean}
    \hat{\alpha}^{(k+1)} = \frac{\overline{|\nabla_{\theta} L_e|}}{ \overline{|\nabla_{\theta} L_b|} },  \quad
    \hat{\beta}^{(k+1)} = \frac{\overline{|\nabla_{\theta} L_e|}}{ \overline{|\nabla_{\theta}L_i|} }.
\end{equation}
The gradients with respect to parameters of the neural network can be easily computed by AD in the deep learning framework. 
Consequently, the weighting coefficients for the next iteration are updated using a moving average form:
\begin{equation}\label{e:dyW_2}
    \alpha^{(k+1)} = (1-\lambda) \alpha^{(k)} + \lambda \hat{\alpha}^{(k+1)},  \quad
    \beta^{(k+1)} = (1-\lambda) \beta^{(k)} + \lambda \hat{\beta}^{(k+1)},
\end{equation}
with $\lambda=0.1$. The strategy of dynamic weights will be applied to most of the NSFnet simulations later. 

We have introduced two different formulations of NSFnets which correspond to the VP form and the VV form. We carry out several numerical experiments with NSFnets of different sizes. However, further accuracy enhancement may be possible for each case presented below using optimization in the size of architecture, the learning rate and even the optimizer, which is beyond the scope of the current work.

\section{Simulations of laminar flows}\label{sec:analyticalExamples}

In this section, we apply the proposed NSFnets to simulate different incompressible Navier-Stokes flows, including 2D steady Kovasznay flow, 2D unsteady cylinder wake and 3D unsteady Beltrami flow. We present comparisons between the VV and VP-NSFnets and investigate the influence of dynamic weights on the accuracy of the solution. Other enhancements can include the use of adaptive activation function to accelerate training \cite{jagtap2020adaptive, jagtap2019locally}, but we did not pursue this in the current work. To evaluate the performance of the NSFnet simulations, we define the relative $L_2$ error at each time step as
\begin{equation}
\epsilon_{\sss V} = \parallel \hat{V} - V \parallel_{2} / \parallel V \parallel_{2},
\label{eq:l2}
\end{equation}
where $V$ denotes the velocity components $(u, v, w)$ or the pressure $p$, and the hat represents the values inferred by NSFnets. The reference velocity and pressure are given by analytical solutions or high-fidelity DNS results. We note that to evaluate the accuracy of NSFnet solutions, we apply a shift for the NSFnet simulation results to bring the means of the pressure for reference DNS results and NSFnet results to the same value.

\subsection{Kovasznay flow}

We use the Kovasznay flow as the first test case to demonstrate the performance of NSFnets. This 2D steady Navier-Stokes flow has the following analytical solution:
\begin{equation}\label{e:Kovasenay}
\begin{aligned}
u(x,y) &= 1-e^{\lambda x} \cos (2\pi y),    \\
v(x,y) &= \frac{\lambda}{2\pi} e^{\lambda x} \sin (2\pi y),   \\
p(x,y) &= \frac{1}{2} (1- e^{2\lambda x}),    
\end{aligned}
\end{equation}
where 
\begin{equation*}
\lambda = \frac{1}{2\nu}- \sqrt[]{\frac{1}{4\nu^{2}}+4\pi^{2}} ,  \quad \nu=\frac{1}{\text{Re}}=\frac{1}{40}. 
\end{equation*}
We consider a computational domain of $[-0.5,1.0]\times[-0.5,1.5]$. There are 101 points with fixed spatial coordinate on each boundary, such that we have 400 training data for the boundary conditions, i.e., $N_{b}=400$. For computing the equation loss of NSFnets, 2601 points are randomly selected inside the domain. There is no initial condition for this steady flow. All the NSFnets are assessed after a two-step training: we first use the \emph{Adam} optimizer for 5,000, 5,000, 50,000 and 50,000 iterations with learning rates of $1\times10^{-3}$, $1\times10^{-4}$, $1\times10^{-5}$ and $1\times10^{-6}$, respectively, then apply the limited-memory Broyden–Fletcher–Goldfarb–Shanno algorithm with bound constraints (L-BFGS-B) to finetune the results. The training process of L-BFGS-B is terminated automatically based on the increment tolerance. 

For Kovasznay flow, we first investigate the influence of the neural network architecture. We employ different sizes of network by varying the number of hidden layers and the number of neurons per layer. The weighting coefficient $\alpha$ for boundary constraint is chosen as 100 for training these NSFnets. The results are summarized in Table \ref{tab:Kovasznay_results_sizes}, where each number is the best over ten independent simulations. As shown in the table, both formulations of NSFnets are able to attain the solutions with high accuracy. The relative errors are in the order of $10^{-5}$ to $10^{-3} $. 
We can also observe that the performance of the NSFnets is improved as the network size increases. The VP formulation outperforms the VV form for small networks, while the VV-NSFnet provides more accurate solutions when using large networks.

\begin{table}[t]
\begin{center}
\caption{Kovasznay flow: relative $L_{2}$ errors of velocity and pressure solutions for NSFnets with different sizes ($\alpha=100$, NN size is the number of hidden layers $\times$ the number of neurons per layer).  }\label{tab:Kovasznay_results_sizes}
\begin{tabular}{c|ccc|cc}
\hline
\multirow{2}{*}{NN size} & \multicolumn{3}{c|}{VP-NSFnet} & \multicolumn{2}{c}{VV-NSFnet} \\ \cline{2-6}
& $\epsilon_{u}$ & $\epsilon_{v}$ & $\epsilon_{p}$ & $\epsilon_{u}$ & $\epsilon_{v}$ \\ \hline
$4\times50$   & 0.076\%  & 0.412\%  & 0.516\%  & 0.131\%    & 0.368\%       \\
$7\times50$   & 0.038\%  & 0.255\%  & 0.114\%  & 0.111\%    & 0.520\%       \\
$7\times100$  & 0.016\%  & 0.183\%  & 0.062\%  & 0.078\%    & 0.397\%       \\
$10\times100$ & 0.020\%  & 0.115\%  & 0.044\%  & 0.040\%    & 0.233\%       \\
$10\times200$ & 0.012\%  & 0.103\%  & 0.042\%  & 0.022\%    & 0.121\%       \\
$10\times250$ & 0.011\%  & 0.101\%  & 0.031\%  & 0.011\%    & 0.081\%       \\
$10\times300$ & 0.008\%  & 0.072\%  & 0.041\%  & 0.004\%    & 0.037\%       \\ \hline
\end{tabular}
\end{center}
\end{table}

\begin{figure}[th]
\begin{center}
\includegraphics[width=9cm]{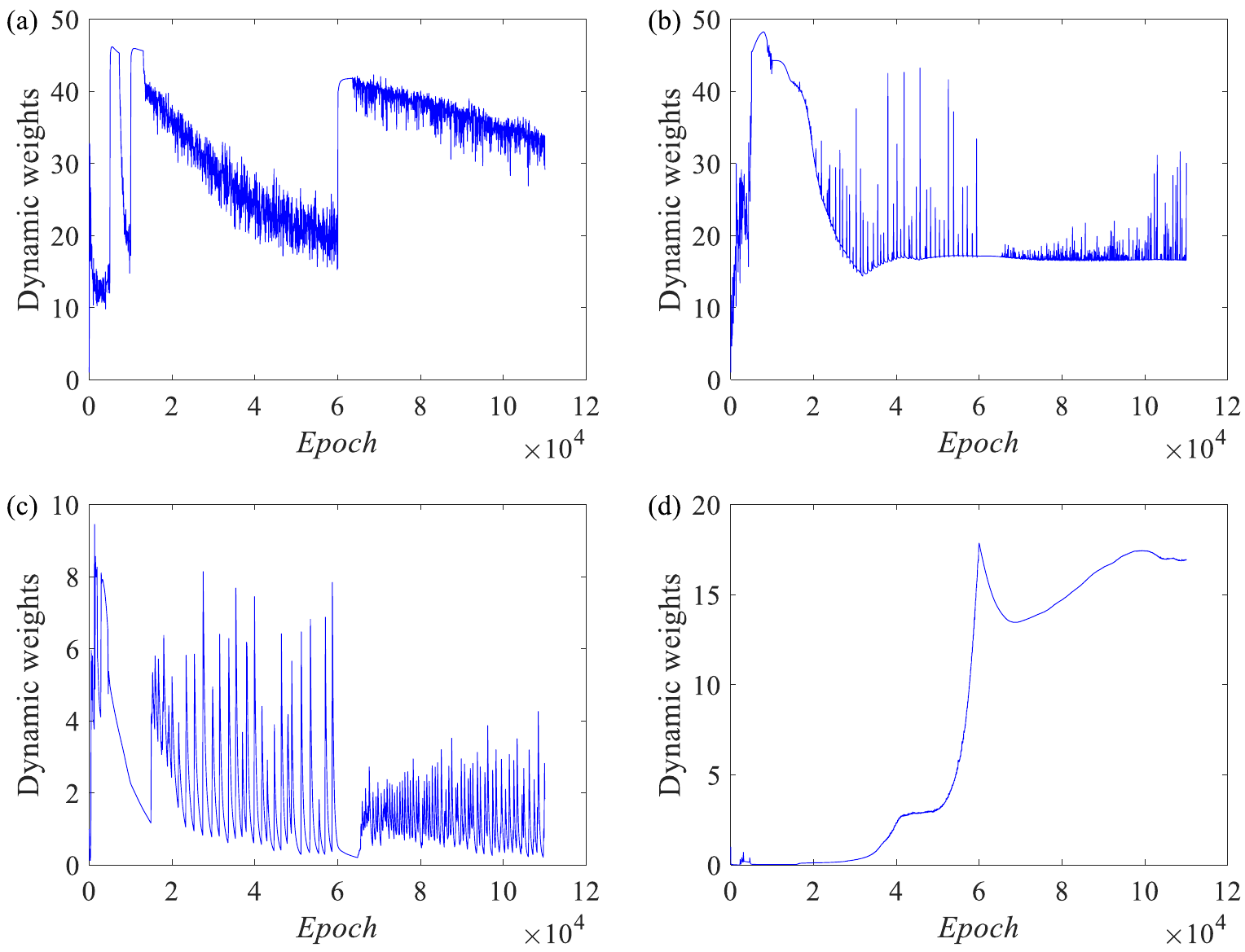}
\caption{Kovasznay flow: dynamic weights for (a) VP-NSFnet, given by Eq. (\ref{e:dyW_1}); (b) VV-NSFnet, given by Eq. (\ref{e:dyW_1}); (c) VP-NSFnet, given by Eq. (\ref{e:dyW_mean}); (d) VV-NSFnet, given by Eq. (\ref{e:dyW_mean}). The NN size is $4\times50$.}\label{fig:Kovasznay_vp_dyW}
\end{center}
\end{figure}

\begin{figure}[th]
\begin{center}
\includegraphics[width=12cm]{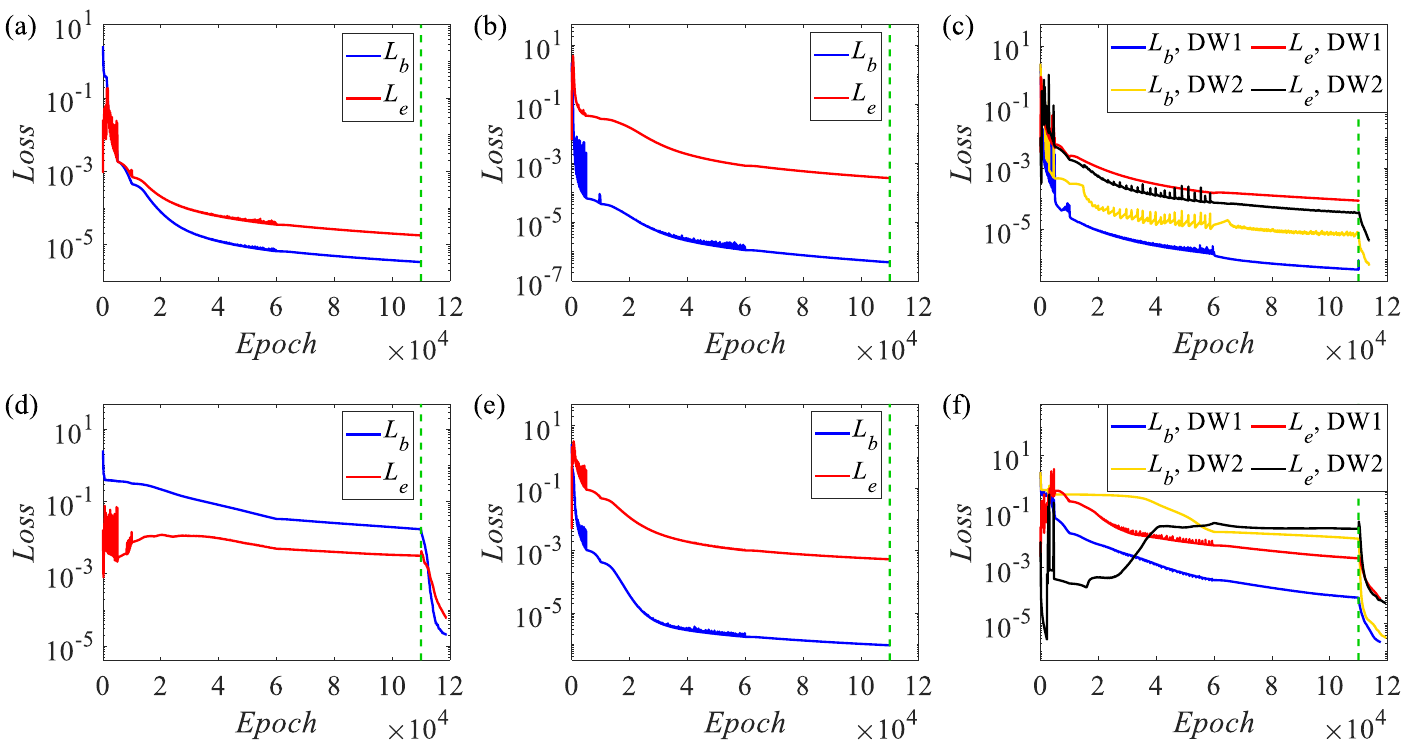}
\caption{Kovasznay flow: loss functions (physics loss $L_e$ and boundary loss $L_b$) obtained by (a) VP-NSFnet, fixed weight $\alpha =1$; (b)  VP-NSFnet, fixed weight $\alpha =100$; (c) VP-NSFnet, dynamic weight; (d) VV-NSFnet, fixed weight $\alpha =1$; (e) VV-NSFnet, fixed weight $\alpha =100$; (f) VV-NSFnet, dynamic weight. ``DW1" denotes dynamic weights given by Eq. (\ref{e:dyW_1}), and ``DW2" denotes dynamic weights given by Eq. (\ref{e:dyW_mean}). \emph{Adam} optimizer is used before the vertical dashed green line, and L-BFGS-B optimizer is used after that. The NN size is $4\times50$. }\label{fig:Kovasznay_vp_loss}
\end{center}
\end{figure}

The weighting coefficient $\alpha$ for the boundary constraint is also investigated here. In addition to letting $\alpha=100$, we also apply $\alpha=1$ and implement dynamic weights (i.e., Eqs. (\ref{e:dyW_1}) and (\ref{e:dyW_mean})) for comparisons. 
In this assessment, we employ a small neural network with 4 hidden layers and 50 neurons per layer. The learning rate is decreasing during the training process as mentioned above. This strategy is consistent with the use of dynamic weights in \cite{wang2020understanding}.
The dynamic weights during the training process are displayed in Fig. \ref{fig:Kovasznay_vp_dyW}. Here $\alpha$ is initialized by 1 for the dynamic weighting strategy. We can find that the coefficient $\alpha$ oscillates and also varies due to the changes of learning rate. 
The dynamic weighting strategy works similarly for VV-NSFnet. The results with dynamic weights are better than those with fixed coefficient value (i.e., $\alpha=1$ and $\alpha=100$). The resulting limit values of $\alpha$ are in the order of $10$. 
The corresponding loss functions obtained by VP-NSFnet with $\alpha=1$, $\alpha=100$ and dynamic weighting strategy are illustrated in Figs. \ref{fig:Kovasznay_vp_loss}(a), (b) and (c), and for the VV-NSFnet in Figs. \ref{fig:Kovasznay_vp_loss}(d), (e) and (f). 
From the curves of the training loss, we find that the \emph{Adam} optimizer is robust for the VP-NSFnet while it does not perform consistently for the VV-NSFnet. Applying two-step optimization can obtain more consistent results. 
The relative $L_2$ errors of the NSFnet simulations with different weights are given in Table \ref{tab:Kovasznay_results}. For fixed weights ($\alpha=1$ and $\alpha=100$), the VP-NSFnet outperforms the VV-NSFnet for simulating Kovasznay flow. The neural network with $\alpha=100$ performs slightly better than that with $\alpha=1$ for both NSFnets. 
When applying dynamic weights during network training, we can obtain more accurate solutions than the former two cases.

\begin{figure}[th]
\begin{center}
\includegraphics[width=9cm]{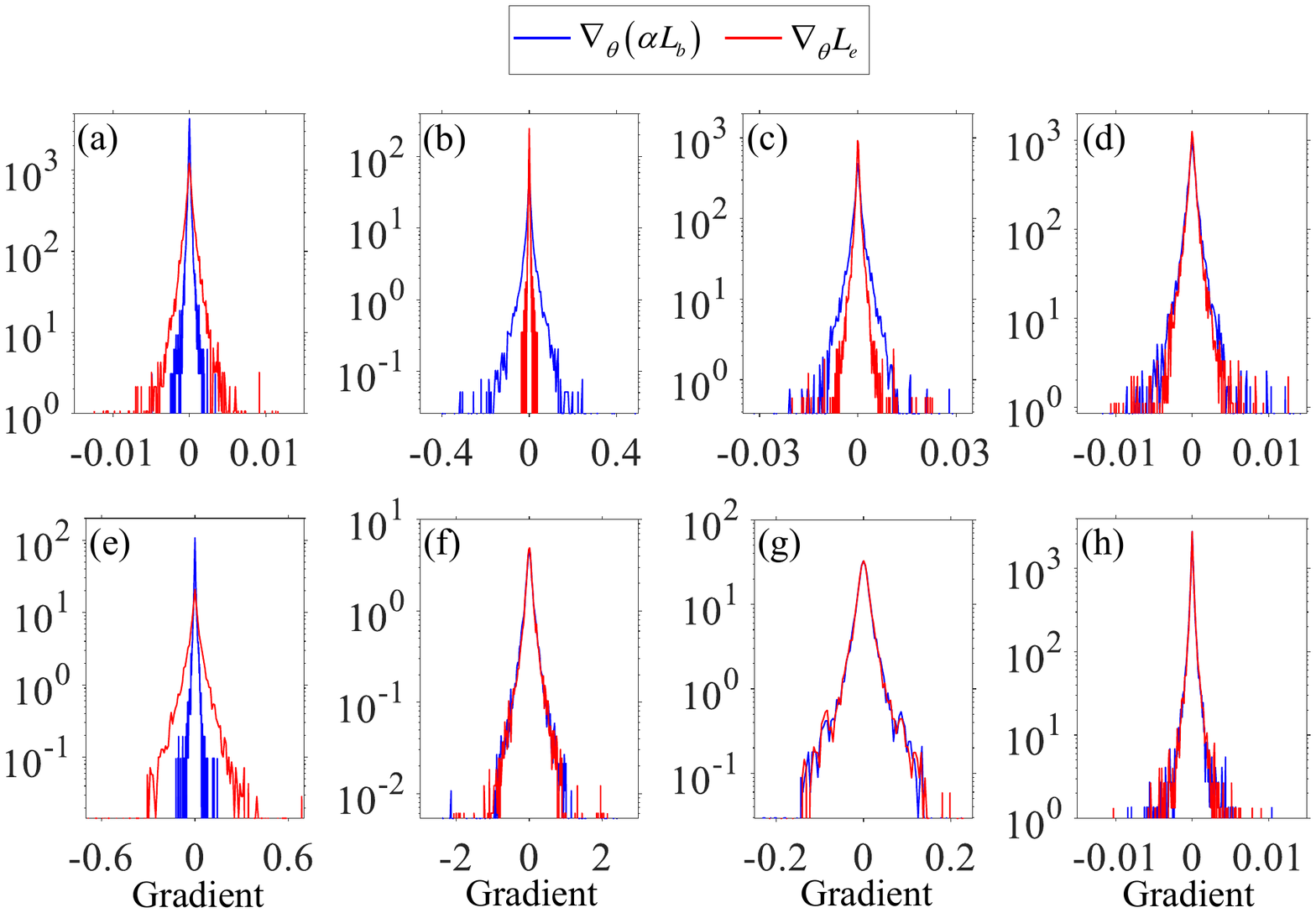}
\caption{Kovasznay flow: histograms of the gradients ($\nabla_{\theta}L_e$ and $\nabla_{\theta} ({\alpha} L_b$)) after 10,000 iterations during training the NSFnets. (a) VP-NSFnet, fixed weight $\alpha =1$; (b)  VP-NSFnet, fixed weight $\alpha =100$; (c) VP-NSFnet, dynamic weight (Eq. (\ref{e:dyW_1})); (d) VP-NSFnet, dynamic weight (Eq. (\ref{e:dyW_mean})); (e) VV-NSFnet, fixed weight $\alpha =1$; (f)  VV-NSFnet, fixed weight $\alpha =100$; (g) VV-NSFnet, dynamic weight (Eq. (\ref{e:dyW_1})); (h) VV-NSFnet, dynamic weight (Eq. (\ref{e:dyW_mean})).  }\label{fig:Kovasznay_gradients}
\end{center}
\end{figure}

In order to demonstrate the effectiveness of dynamic weights, we analyze the gradients of the loss function with respect to the parameters of the NSFnets. The histograms of the back-propagated gradients ($\nabla_{\theta}L_e$ and $\nabla_{\theta} ({\alpha} L_b)$) after 10,000 iterations are shown in Fig. \ref{fig:Kovasznay_gradients}. Our goal is to balance the distributions of $\nabla_{\theta}L_e$ and $\nabla_{\theta} ({\alpha} L_b)$, thus these two terms can contribute equally to the parameter updating (i.e., Eq. (\ref{e:theta_iterations})). As shown in Figs. \ref{fig:Kovasznay_gradients}(a) and \ref{fig:Kovasznay_gradients}(e), the gradients of two different terms are unbalanced when there is no weighting coefficient for the boundary constraints (i.e., $\alpha=1$). When applying dynamic weights, the histograms of $\nabla_{\theta}L_e$ and $\nabla_{\theta} ({\alpha} L_b)$ are more consistent with each other. For the VP-NSFnet, the second formulation of dynamic weights, i.e., Eq. (\ref{e:dyW_mean}), performs better than the first one (weights given by Eq. (\ref{e:dyW_1})). However, the VV-NSFnet behaves in the opposite way for this case, as shown in Figs. \ref{fig:Kovasznay_gradients}(g) and \ref{fig:Kovasznay_gradients}(h).

\begin{table}[t]
\begin{center}
\caption{Kovasznay flow: relative $L_{2}$ errors of velocity and pressure solutions with different weights. The NN size is $4\times50$.}\label{tab:Kovasznay_results}
\begin{tabular}{c|ccc|cc}
\hline
\multirow{2}{*}{Weights} & \multicolumn{3}{c|}{VP-NSFnet} & \multicolumn{2}{c}{VV-NSFnet} \\ \cline{2-6}
& $\epsilon_{u}$ & $\epsilon_{v}$ & $\epsilon_{p}$ & $\epsilon_{u}$ & $\epsilon_{v}$ \\ \hline
$\alpha = 1$     & 0.084\%  & 0.425\%  & 0.309\%  & 0.211\%    & 1.071\%       \\
$\alpha = 100$   & 0.076\%  & 0.412\%  & 0.516\%  & 0.131\%    & 0.368\%       \\
Dynamic, Eq. (\ref{e:dyW_1})      & 0.072\%  & 0.352\%  & 0.212\%  & 0.056\%    & 0.436\%       \\
Dynamic, Eq. (\ref{e:dyW_mean})      & 0.026\%  & 0.199\%  & 0.141\%  & 0.067\%    & 0.446\%       \\
\hline
\end{tabular}
\end{center}
\end{table}

\subsection{Two-dimensional cylinder wake}
Here we use NSFnets to simulate the 2D vortex shedding behind a circular cylinder at $\text{Re}=100$. The cylinder is placed at $(x,y)=(0,0)$ with diameter $D=1$. High-fidelity DNS data from \citep{raissi2019physics} is used as a reference and for providing boundary and initial data for NSFnet training. We consider a domain defined by $[1,8]\times [-2,2]$ and the time interval is $[0,7]$ (about one shedding period) with time step $\Delta t =0.1$. As for the training data, we place 100 points along the the $x$-direction boundary and 50 points along the $y$-direction boundary to enforce the boundary conditions and use 140,000 spatio-temporal scattered points inside the domain to compute the residuals. The NSFnets contain 10 hidden layer and 100 neurons per layer. 
In addition to the default models with $\alpha=\beta=1$ and $\alpha=\beta=100$, we again implement the dynamic weighting strategy for NSFnets. 
The training procedure is identical to the one we used for Kovasznay flow. 
%

\begin{figure}[th]
\begin{center}
\includegraphics[width=12cm]{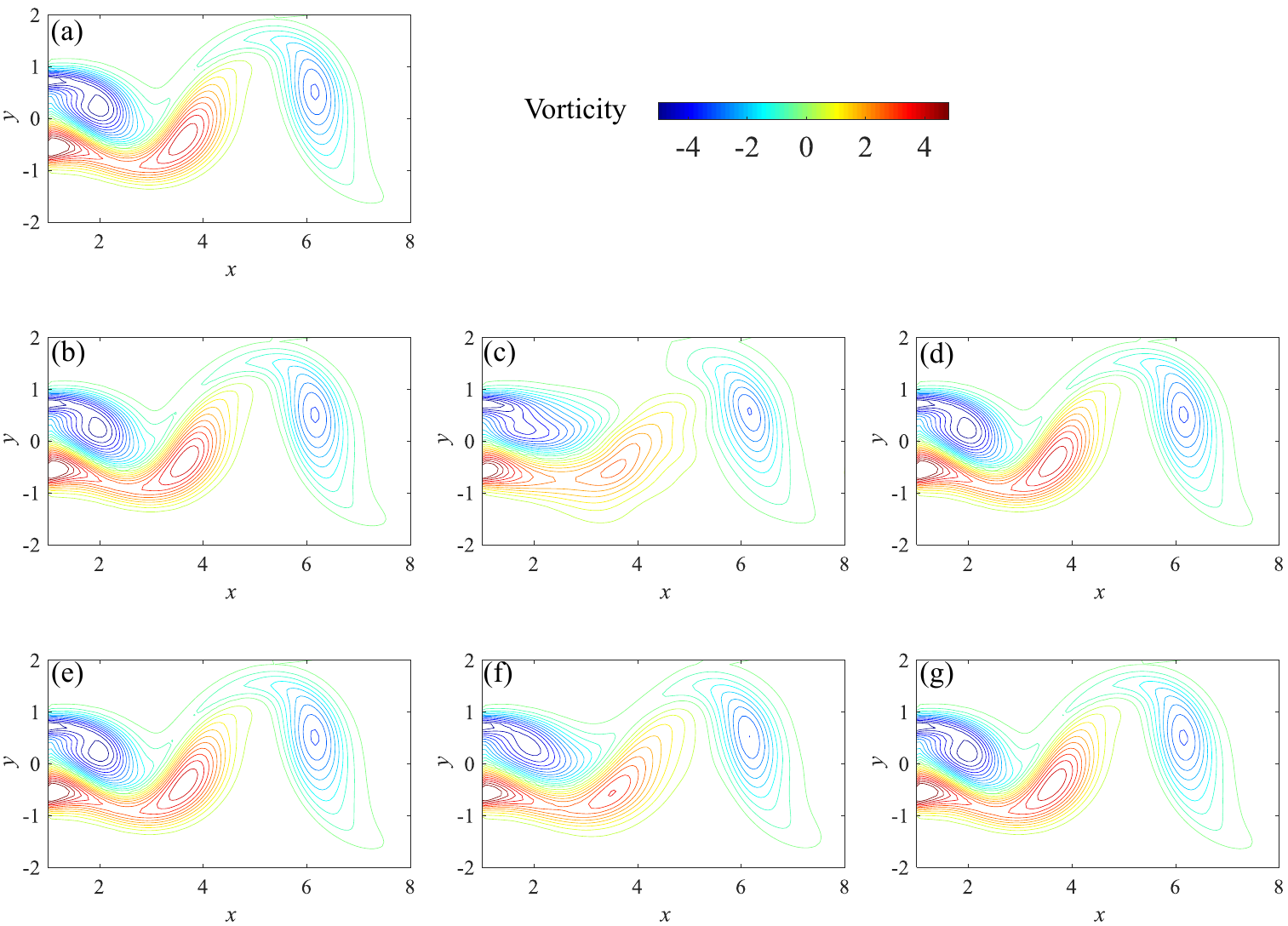}
\caption{Flow past a circular cylinder: contours of the vorticity on the same contour levels at $t=4.0$: (a) reference DNS solution from \citep{raissi2019physics}; (b) VP-NSFnet, fixed weights $\alpha =\beta =1$; (c) VP-NSFnet, fixed weights $\alpha =\beta =100$; (d )VP-NSFnet, dynamic weights; (e) VV-NSFnet, fixed weights $\alpha =\beta =1$; (f) VV-NSFnet, fixed weights $\alpha =\beta =100$; (g) VV-NSFnet, dynamic weights. The dynamic weights here are given by Eq. (\ref{e:dyW_1}).}
\label{fig:cylinderContour}
\end{center}
\end{figure}

\begin{figure}[th]
\begin{center}
\includegraphics[width=12cm]{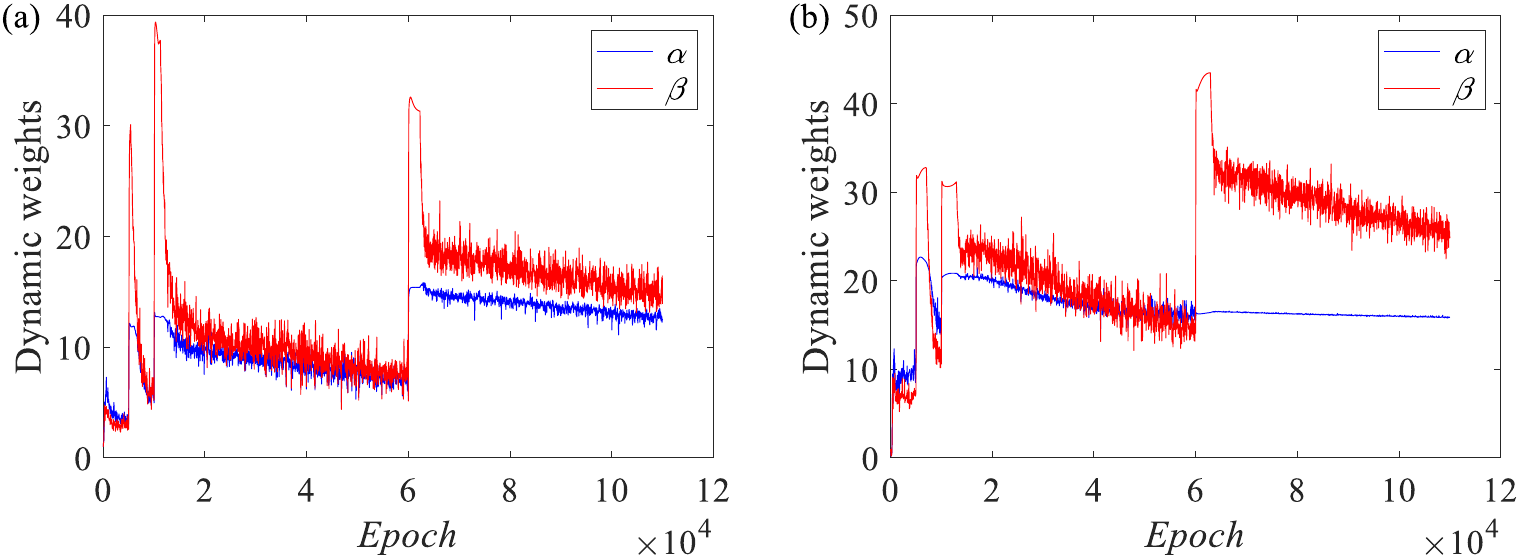}
\caption{Flow past a circular cylinder: dynamic weights  given by Eq. (\ref{e:dyW_1}) for (a) VP-NSFnet and (b) VV-NSFnet. }\label{fig:cylinder_dyW}
\end{center}
\end{figure}

\begin{figure}[th]
\begin{center}
\includegraphics[width=12cm]{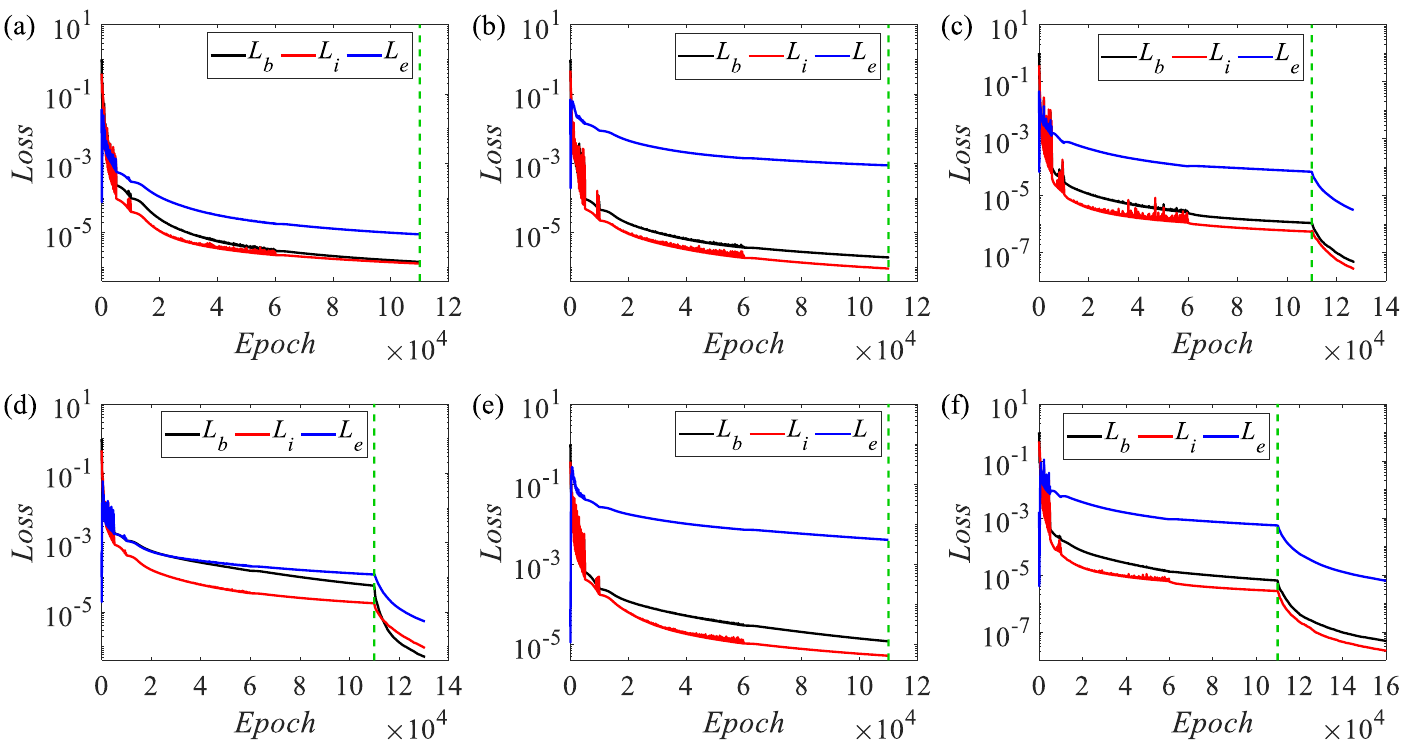}
\caption{Flow past a circular cylinder: loss functions (physics loss $L_e$ and boundary loss $L_b$) obtained by (a) VP-NSFnet, fixed weights $\alpha=\beta=1$; (b) VP-NSFnet, fixed weights $\alpha=\beta=100$; (c) VP-NSFnet, dynamic weights given by Eq. (\ref{e:dyW_1}); (d) VV-NSFnet, fixed weights $\alpha =\beta=1$; (e) VV-NSFnet, fixed weights $\alpha =\beta=100$; (f) VV-NSFnet, dynamic weights given by Eq. (\ref{e:dyW_1}). \emph{Adam} optimizer is used before the dashed green line, and L-BFGS-B optimizer is used after the dashed green line. The NN size is $10\times100$. }\label{fig:cylinderLoss}
\end{center}
\end{figure}

\begin{figure}[th]
\begin{center}
\includegraphics[width=12cm]{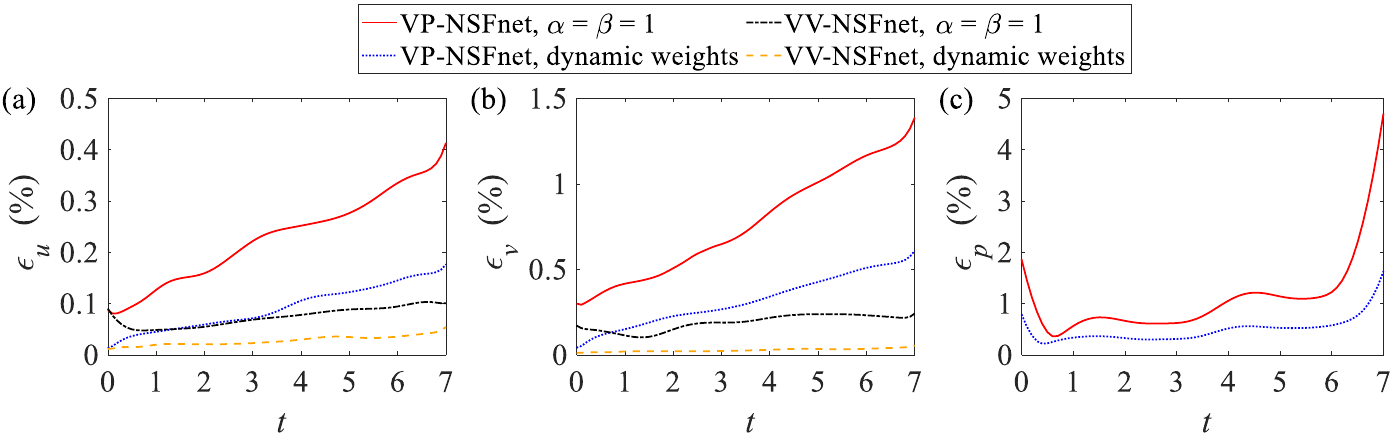}
\caption{Flow past a circular cylinder: relative $L_{2}$ errors of NSFnets simulations for (a) the streamwise velocity, (b) the crossflow velocity and (c) pressure. The dynamic weights here are given by Eq. (\ref{e:dyW_1}).}\label{fig:cylinder_errors}
\end{center}
\end{figure}

A snapshot of the vorticity contours at $t=4.0$ is shown in Fig. \ref{fig:cylinderContour}, demonstrating qualitative agreement of NSFnet inference with the DNS result. 
In Fig. \ref{fig:cylinder_dyW} we present the dynamic weights of both VP- and VV-NSFnets. In this case, the weights are both initialized by the value 1. The variations of the weights correspond to the changes of learning rates. We can observe that both $\alpha$ and $\beta$ are in the order of 10 and the weights for initial conditions are larger than the weights for boundary conditions in this case. The separated terms of the weighted loss function during training are given in Fig. \ref{fig:cylinderLoss}. We employ two-step training to ensure the convergence for all the NSFnets. The relative $L_{2}$ errors of NSFnet simulations versus time are given in Fig. \ref{fig:cylinder_errors}. We see that the VV-NSFnet performs better than the VP-NSFnet, and also that applying the dynamic weights can improve the simulation accuracy for both formulations.

\subsection{Three-dimensional Beltrami flow}

\begin{figure}[th]
\begin{center}
\includegraphics[width=12cm]{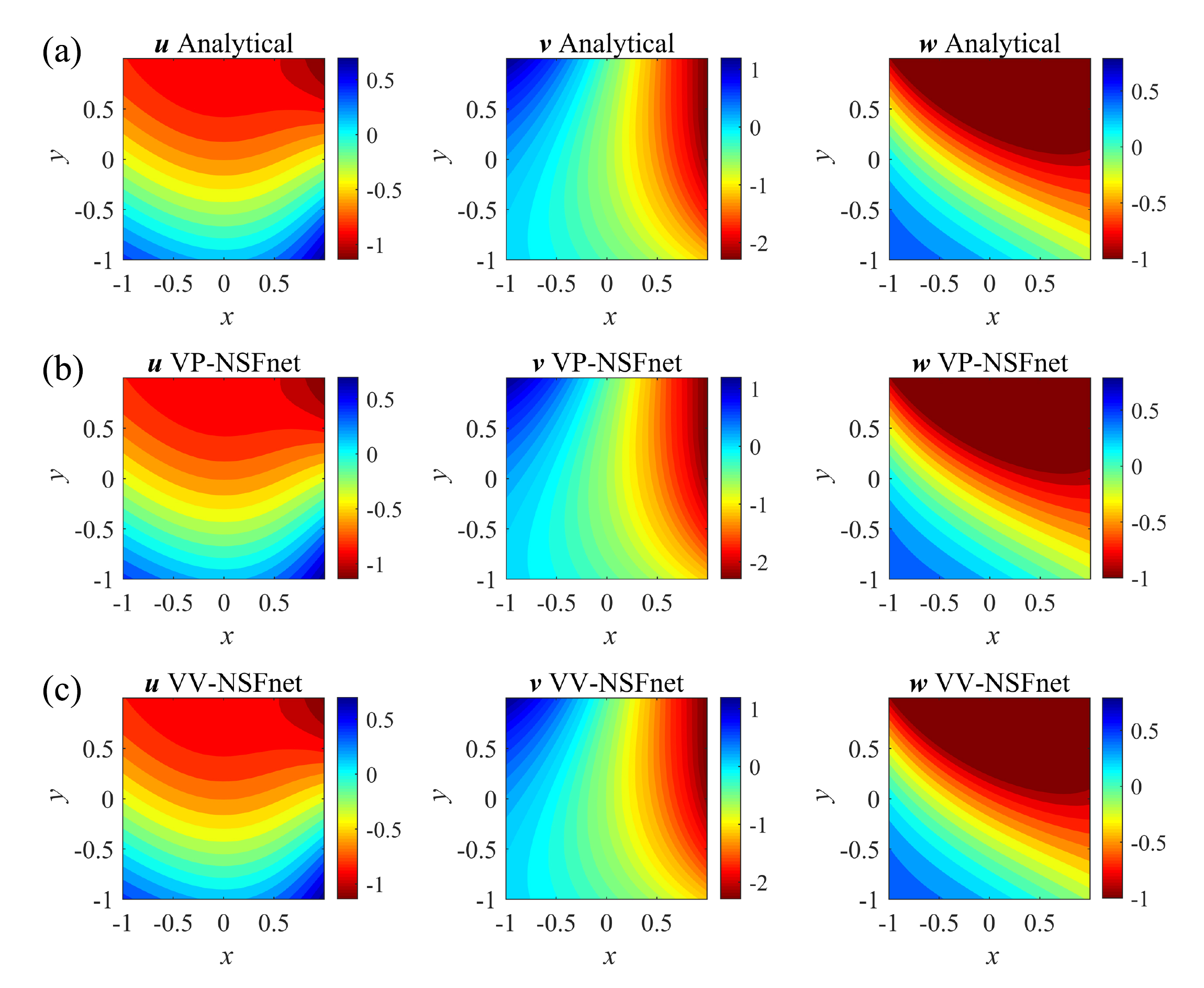}
\caption{Beltrami flow: velocity fields at $t=1$ on the plane $z=0$. (a) analytical solutions; 
(b) results of VP-NSFnet; (c) results of VV-NSFnet. }
\label{fig:Beltrami_fields_t1_z0}
\end{center}
\end{figure}

The analytical solutions of the unsteady three-dimensional Beltrami flow developed by \citet{ethier1994exact} are:
\begin{equation}\label{e:Beltrami}
\begin{aligned}
u(x,y,z,t) =& -a \left[ e^{ax}\sin(ay+dz) + e^{az}\cos(ax+dy)  \right] e^{-d^{2}t},    \\
v(x,y,z,t) =& -a \left[ e^{ay}\sin(az+dx) + e^{ax}\cos(ay+dz)  \right] e^{-d^{2}t},   \\
w(x,y,z,t) =& -a \left[ e^{az}\sin(ax+dy) + e^{ay}\cos(az+dx)  \right] e^{-d^{2}t},
\\
p(x,y,z,t) =& -\frac{1}{2}a^{2} \left[ e^{2ax} + e^{2ay} + e^{2az} +2\sin(ax+dy)\cos(az+dx)e^{a(y+z)} \right.   \\ 
& +2\sin(ay+dz)\cos(ax+dy)e^{a(z+x)}  \\
& \left. +2\sin(az+dx)\cos(ay+dz)e^{a(x+y)} \right] e^{-2d^{2}t},
\end{aligned}
\end{equation}
where $a=d=1$ is used. 
In NSFnet simulations, the computational domain is defined by $[-1,1] \times [-1,1] \times [-1,1]$ and the time interval is $[0,1]$; the time step is 0.1. 
For the NSFnet training data, $31\times31$ points on each face are used for boundary conditions while a batch of 10,000 points in the spatio-temporal domain is used for the equations. The weighting coefficients of the loss function are both fixed in this case: $\alpha=\beta=100$. 
The two-step optimization (\emph{Adam} and L-BFGS-B) is implemented to train the neural networks, which have the default architecture with 10 layers and 100 neurons per layer. 
A snapshot of the velocity fields at $t=1$ and on the slice $z=0$ is displayed in Fig. \ref{fig:Beltrami_fields_t1_z0}.
The errors of simulation results of the two NSFnets at different time steps are presented in Table \ref{tab:Beltrami_results}, where  
the relative $L_{2}$ errors of the three velocity components are given. As shown, both NSFnets can obtain accurate solutions of 
the Navier-Stokes equations for the Beltrami flow, but the VV-NSFnet outperforms the VP-NSFnet.

\begin{table}[th]
\begin{center}
\caption{Beltrami flow: relative $L_{2}$ errors for VP-NSFnet and VV-NSFnet.  }\label{tab:Beltrami_results}
\begin{tabular}{c|cccc|ccc}
\hline
\multirow{2}{*}{$t$} & \multicolumn{4}{c|}{VP-NSFnet} & \multicolumn{3}{c}{VV-NSFnet} \\ \cline{2-8}
& $\epsilon_{u}$ & $\epsilon_{v}$ & $\epsilon_{w}$ & $\epsilon_{p}$ & $\epsilon_{u}$ & $\epsilon_{v}$ & $\epsilon_{w}$\\ \hline
$0$    & 0.067\%  & 0.059\%  &0.061\%& 0.700\%  & 0.069\%    & 0.067\%  & 0.066\%     \\
$0.25$ & 0.158\%  & 0.132\%  &0.140\%& 0.778\%  & 0.109\%    & 0.094\%  & 0.108\%    \\
$0.50$ & 0.221\%  & 0.189\%  &0.233\%& 1.292\%  & 0.118\%    & 0.119\%  & 0.132\%    \\
$0.75$ & 0.287\%  & 0.217\%  &0.406\%& 2.149\%  & 0.156\%    & 0.154\%  & 0.187\%    \\
$1.00$ & 0.426\%  & 0.366\%  &0.587\%& 4.766\%  & 0.255\%    & 0.284\%  & 0.263\%    \\ \hline
\end{tabular}
\end{center}
\end{table}
\clearpage

\section{Simulations of turbulent channel flow}\label{sec:turbulentExamples}
\subsection{Problem setup}
We simulate turbulent channel flow at $ \rm Re_\tau = 9.9935 \times 10^2$ systematically by using VP-NSFnets. We use the turbulent channel flow database \citep{perlman2007data,li2008public,graham2016web} at \url{http://turbulence.pha.jhu.edu} as the reference DNS solution. The database provides both the reference and some initial or boundary conditions for the VP-NSFnet. The DNS domain for the channel flow in the database is [0, 8$\pi$] $ \times $ [-1, 1] $ \times $ [0, 3$\pi$]; the mean pressure gradient is $dP/dx = 0.0025$. 
The non-dimensional time step for DNS is 0.0013 while the online database time step is 0.0065 (five times of that for DNS). So, the time step of 0.0065 is also used for evaluating the residuals in NSFnets, i.e., five times of that for DNS. We perform NSFnet simulations by considering different subdomains with different sizes at various locations in the channel. In the first example, we place a box with size $ \sim $200 in wall-units covering a long time period. Then, we test the NSFnet simulation at a larger domain covering half the channel height. Finally, we check the influence of hyperparameters to the NSFnet simulation accuracy. We employ mini-batch to train NSFnets in this study. There are three parts in the input data corresponding to the initial conditions, the boundary conditions and the residuals of equations, respectively. Therefore, we specify the total number of iterations $n_{it}$ in one training epoch, and data in each part are evenly divided into $n_{it}$ small mini-batches. The total data in an entire mini-batch include data from each small mini-batch.

\subsection{Simulation results over a long time interval}
\begin{figure}
	\centerline{\includegraphics [width=12cm] {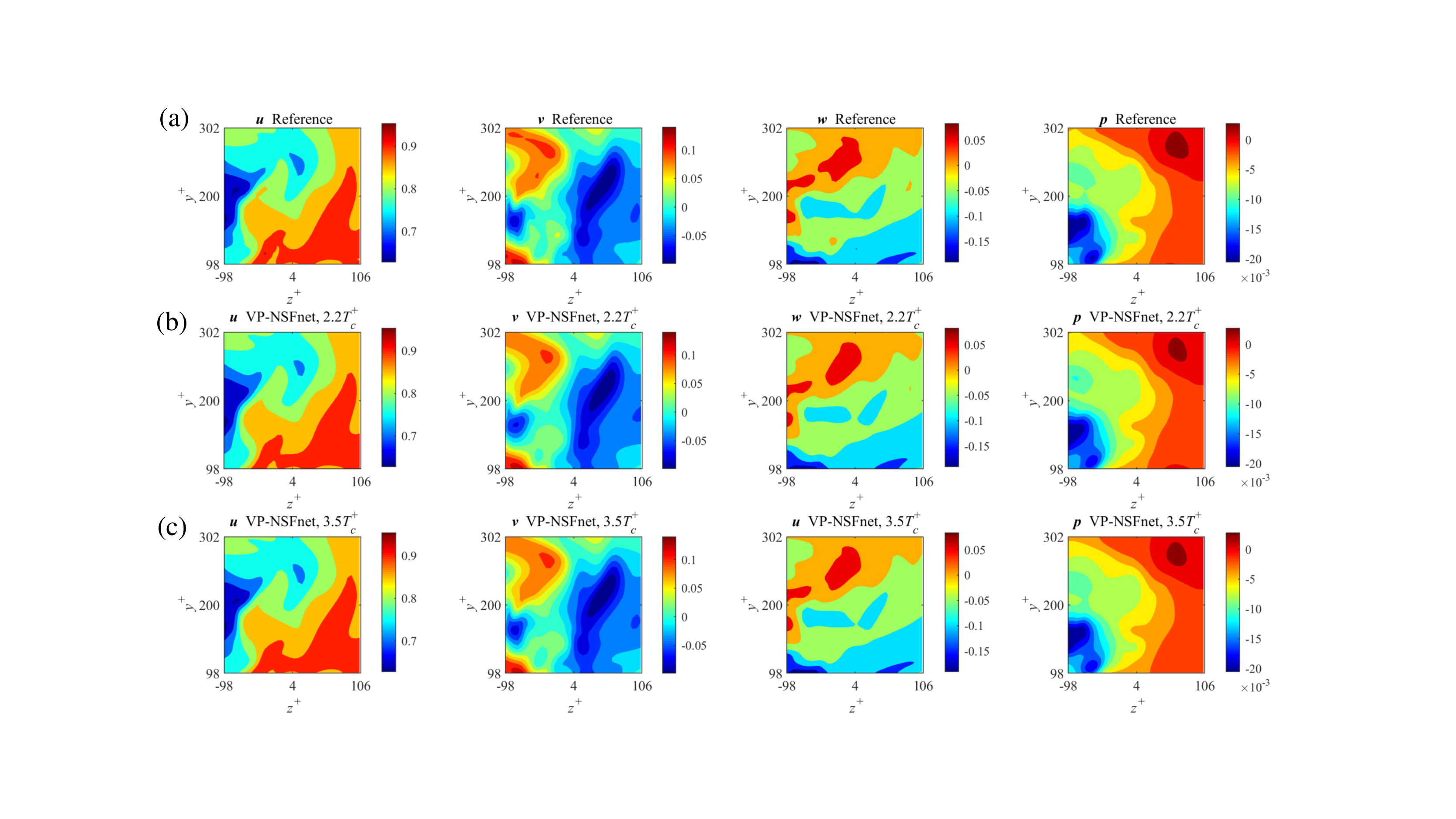}}
	\caption{Long time interval: comparisons of instantaneous $ z-y $ plane flow fields between reference DNS and VP-NSFnet at $ t^+ $ = 24.67: (a) reference solutions; (b) VP-NSFnet, simulation covers $2.2T_c^+$; (c) VP-NSFnet, simulation covers $3.5T_c^+$.}
	\label{fig:contourLong}
\end{figure}

\begin{figure}
	\centerline{\includegraphics [width=12cm] {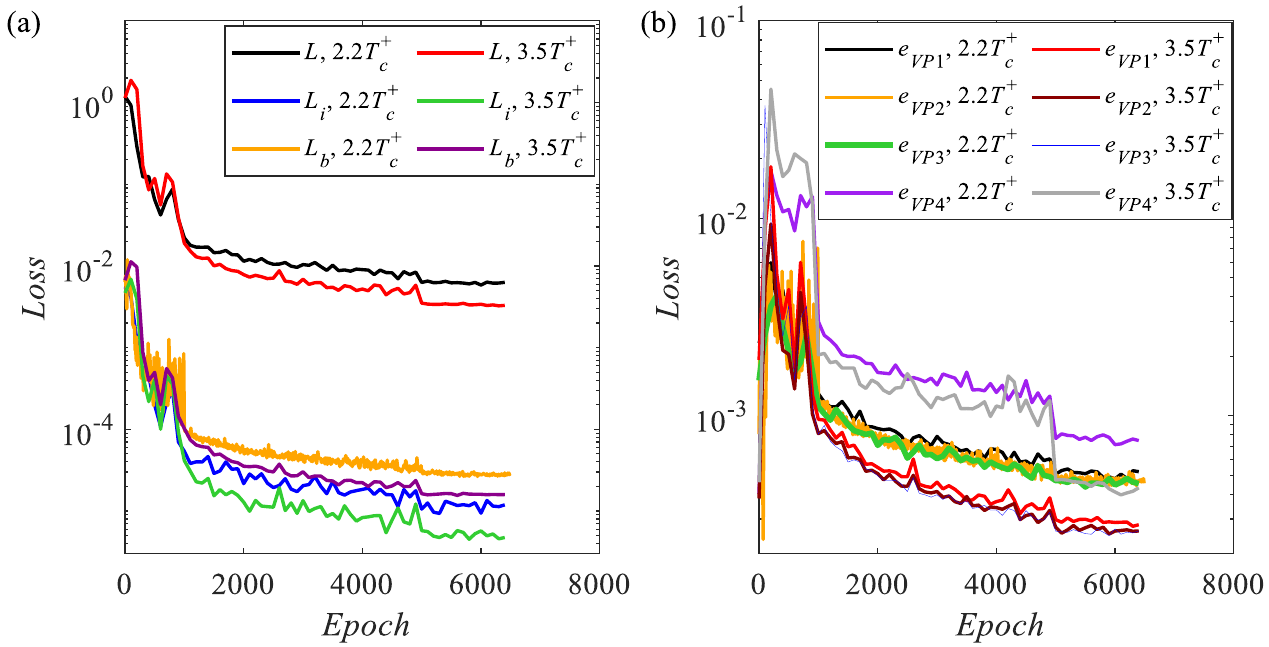}}
	\caption{Long time interval: convergence of the VP-NSFnet simulation: (a) convergence of total loss functions and boundary loss functions; (b) convergence of residuals of governing equations.}
	\label{fig:lossLong}
\end{figure}

\begin{figure}
	\centerline{\includegraphics [width=12cm] {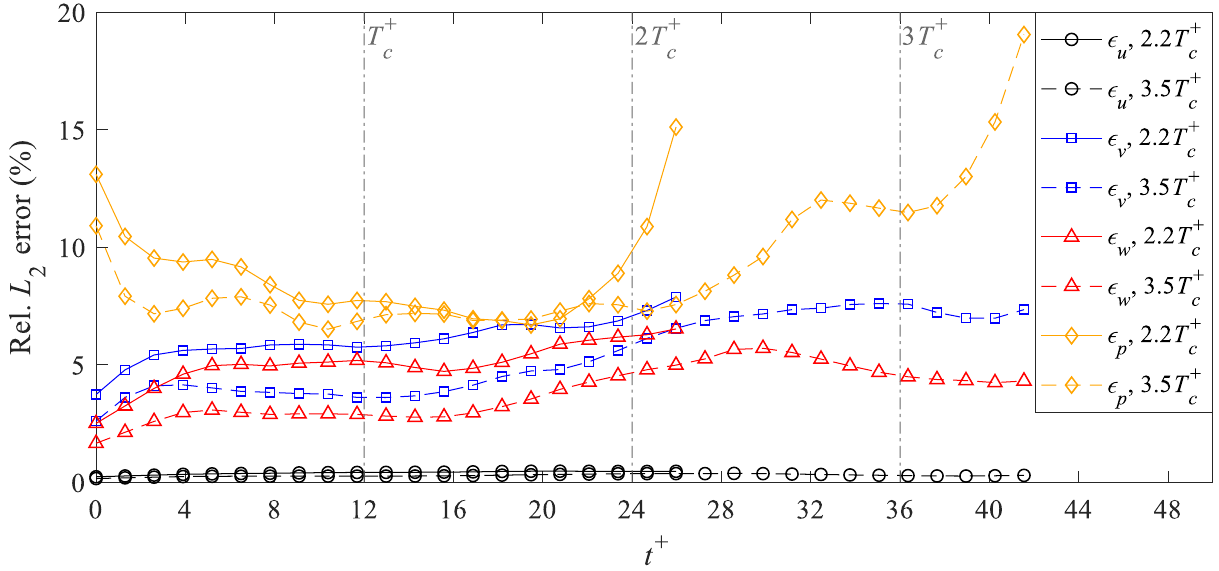}}
	\caption{Long time interval: accuracy of the VP-NSFnet simulation: the solid lines and dashed lines denote simulations covering $2.2T_c^+$ and $3.5T_c^+$, respectively.}
	\label{fig:errLong}
\end{figure}

We first investigate if the VP-NSFnet can sustain turbulence, so  we carry out simulations covering a relatively long time interval. In this test, a subdomain at [12.47, 12.66] $ \times $ [-0.90, -0.70] $ \times $ [4.61, 4.82] (190 $ \times $ 200 $ \times $ 210 in wall-units) is considered as the simulation domain of VP-NSFnet. We perform two different simulations covering the non-dimensional time domain [0, 0.52] (81 time steps, 25.97 in wall-units) and [0, 0.832] (129 time steps, 41.55 in wall-units). Here, we define the local convective time unit of the simulation region as $ T_c^ +  = {L_x^+}/U(y)_{min} = 12.0 $. ($L_x^+$ is the size of the domain in streamwise direction.) Therefore, 25.97 and 41.55 cover more than two convective time units, i.e., $2.2T_c^+$ and $3.5T_c^+$ respectively, for this example. We use 20,000 points inside the domain, 6,644 points on the boundary sampled at each time step, together with 33,524 points at the initial time step to compute the loss function. We set the total number of iterations $n_{it} = 150$ in one training epoch. There are 10 hidden layers in the VP-NSFnet with 300 neurons per layer. The initial learning rate for \emph{Adam} decays from $ 10^{-3} $ (1,000 training epochs) to $ 10^{-4} $ (4,000 training epochs), $ 10^{-5} $ (1,000 training epochs) and $ 10^{-6} $ (500 training epochs) in the training process. The weights in Eq. (\ref{e:loss_vp_L1}) are $ \alpha $ = 100, $ \beta $ = 100. The comparisons of instantaneous flow fields between reference DNS and VP-NSFnet at $ t^+ $ = 24.67 are given in Fig. \ref{fig:contourLong}. The convergence of loss functions is shown in Fig. \ref{fig:lossLong}. The comparisons for accuracy of VP-NSFnet solutions are shown in Fig. \ref{fig:errLong}. All the simulation errors of velocity components are less than 10\%, but the relative $L_2$ error of pressure can reach 15\% or 19\%. Overall, a good VP-NSFnet simulation accuracy is obtained. These results suggest that VP-NSFnet can sustain turbulence for a long time period.

\subsection{Simulation results over a large domain}

\begin{figure}
	\centerline{\includegraphics [width=12cm] {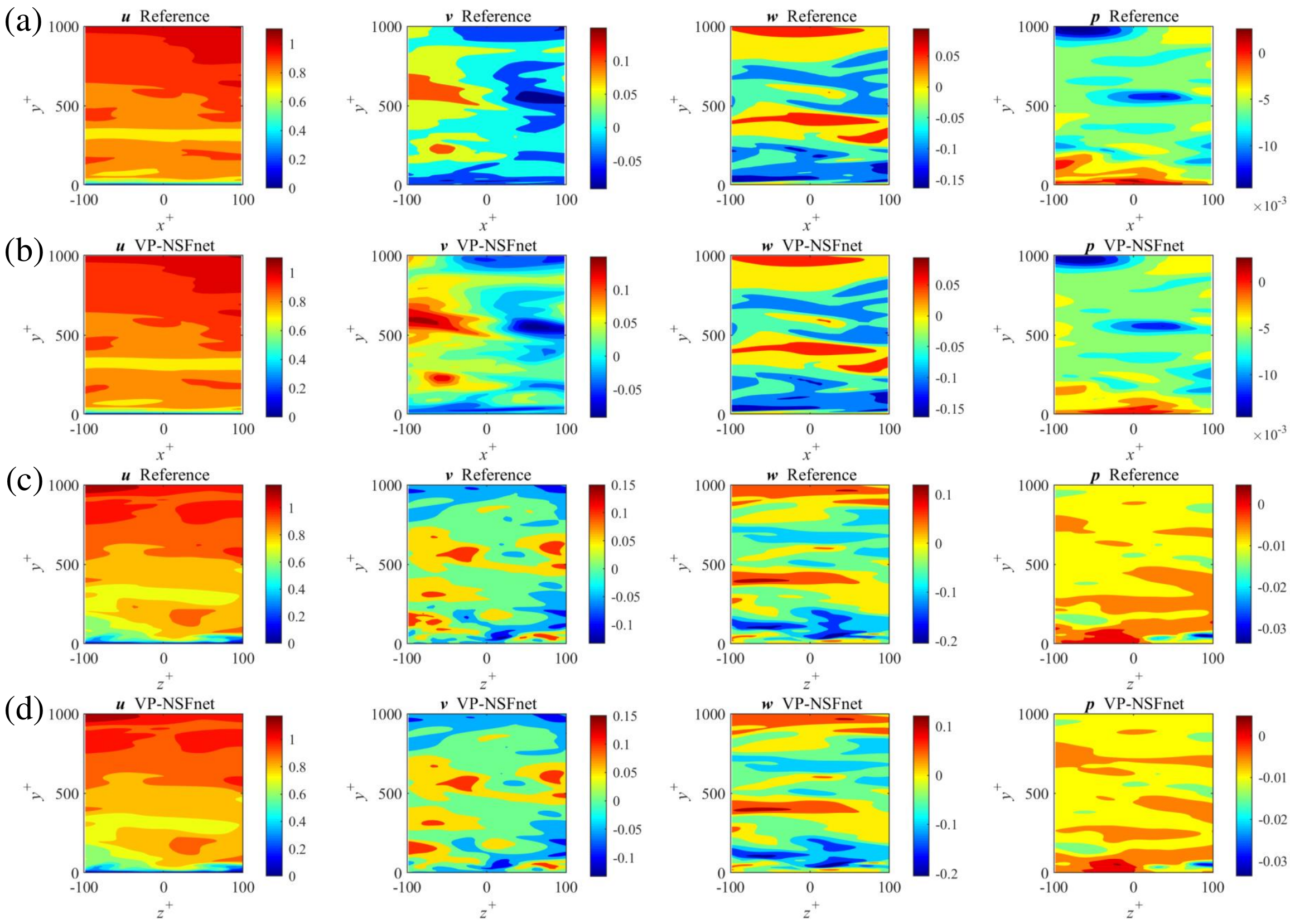}}
	\caption{Large domain: comparisons of instantaneous flow fields between reference DNS and VP-NSFnet at $ t^+ $ = 5.19 covering half channel height: (a) reference solutions, $ x-y $ plane, $z^+ = 0$; (b) VP-NSFnet, $ x-y $ plane, $z^+ = 0$; (c) reference solutions, $ z-y $ plane, $x^+ = 0$; (d) VP-NSFnet, $ z-y $ plane, $x^+ = 0$.}
	\label{fig:contour1000}
\end{figure}

\begin{figure}
	\centerline{\includegraphics [width=12cm] {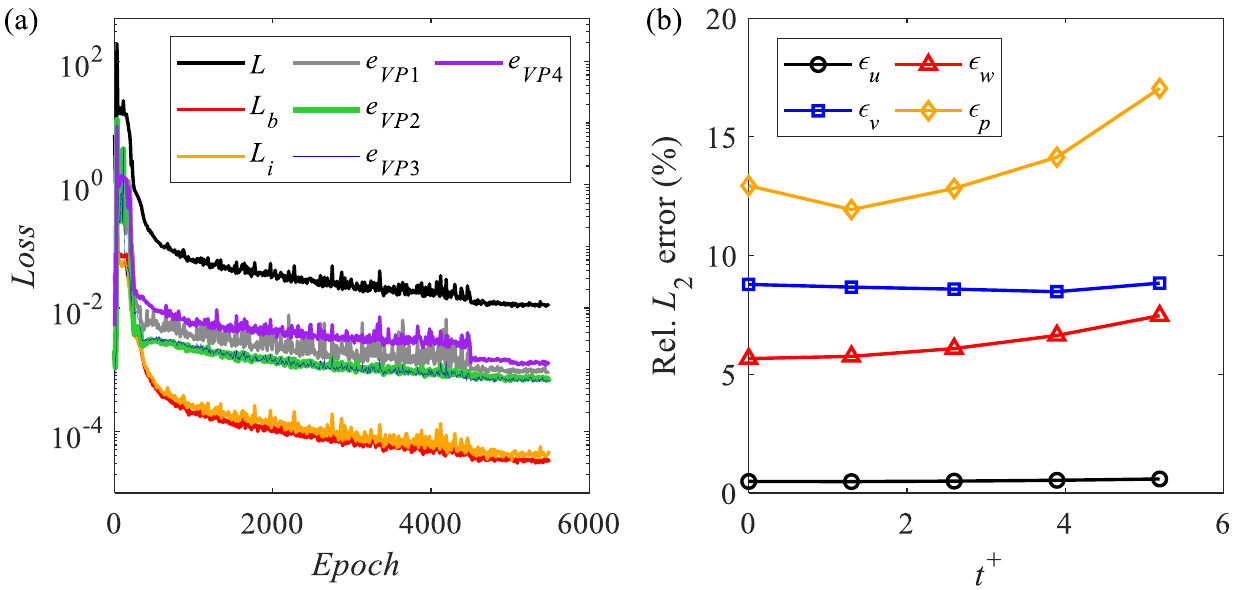}}
	\caption{Large domain: convergence and accuracy of VP-NSFnet covering half channel height: (a) convergence of loss function; (b) relative $L_2$ error.}
	\label{fig:err1000}
\end{figure}

We consider a larger domain covering half channel height in this test. We consider a subdomain at [12.47, 12.66] $ \times $ [-1, -0.0031] $ \times $ [4.61, 4.82] (about 190 $ \times $ 997 $ \times $ 210 in wall-units) as the VP-NSFnet simulation domain; and the non-dimensional time domain is set as [0, 0.104] (17 time steps, 5.19 in wall-units). We place 100,000 points inside the domain and 26,048 points on the boundary sampled at each time step, and 147,968 points at the initial time step to determine the loss function. The total number of iterations $n_{it}$ in one training epoch is taken as 150. There are 10 hidden layers in the VP-NSFnet, with 300 neurons per layer. The initial learning rate for \emph{Adam} decays from $ 10^{-3} $ (250 training epochs) to $ 10^{-4} $ (4,250 training epochs), $ 10^{-5} $ (500 training epochs) and $ 10^{-6} $ (500 training epochs) in this numerical example. The weights in Eq.  (\ref{e:loss_vp_L1}) are $ \alpha $ = 100, $ \beta $ = 100. Comparisons of instantaneous flow fields between reference DNS and VP-NSFnet at the final simulation time step are shown in Fig. \ref{fig:contour1000}. The convergence and accuracy of VP-NSFnet solutions are shown in Fig. \ref{fig:err1000}. All the simulation errors of velocity components are less than 10\%, but the relative $L_2$ error of pressure can reach 17\%. In such a large domain, complex interactions between different scales of eddies occur, and this domain covers the whole range including the law of the wall, the viscous sub-layer, the buffer layer, the log-law region and the outer layer \citep{pope2000turbulent}. However, VP-NSFnet can still get very accurate solutions. The results also indicate that the relative $L_2$ errors of the wall-normal and spanwise velocities are much higher than that of streamwise velocity, i.e., nearly an order higher. This is caused by larger amplitude of the streamwise velocity than the other two velocity components, also nearly an order higher, as shown in Fig. \ref{fig:err1000}. A proper normalization with careful tuning of anisotropic weights may result in a more balanced accuracy.

\subsection{Influence of weights}

\begin{figure}
	\centerline{\includegraphics [width=12cm] {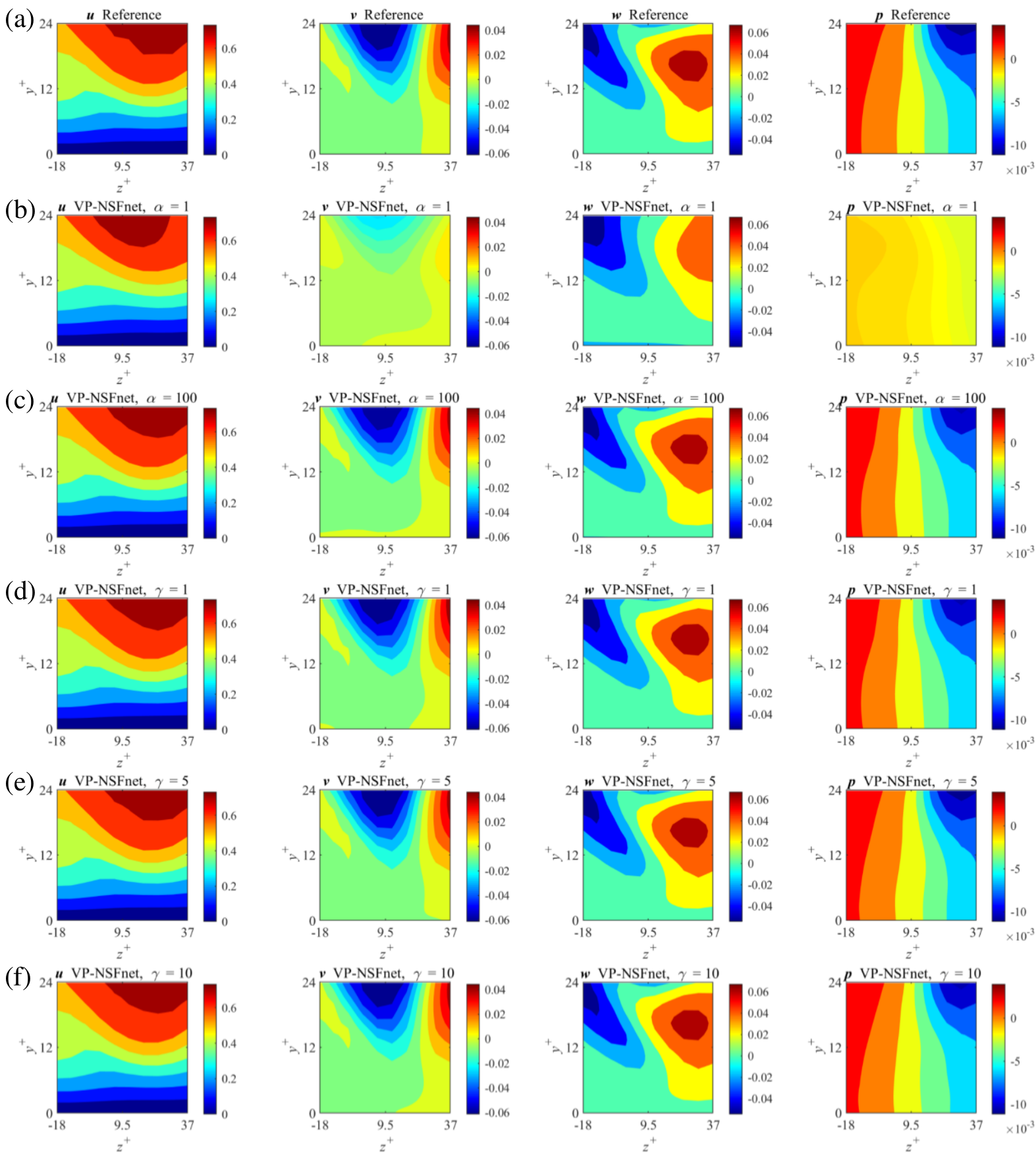}}
	\caption{Influence of weights: comparisons of instantaneous flow fields ($ z-y $ plane) between reference DNS and VP-NSFnet at $ t^+ $ = 5.19 and $x^+ = -12.27$: (a) reference solutions; (b) VP-NSFnet, fixed weight $\alpha = 1$; (c) VP-NSFnet, fixed weight $\alpha = 100$; (d) VP-NSFnet, dynamic weight with normalization factor $\gamma = 1$; (e) VP-NSFnet, dynamic weight with normalization factor $\gamma = 5$; (f) VP-NSFnet, dynamic weight with normalization factor $\gamma = 10$. All the dynamic weights here are given by Eq. (\ref{e:dyW_dns}).}
	\label{fig:hyperContour}
\end{figure}

\begin{figure}
	\centerline{\includegraphics [width=6cm] {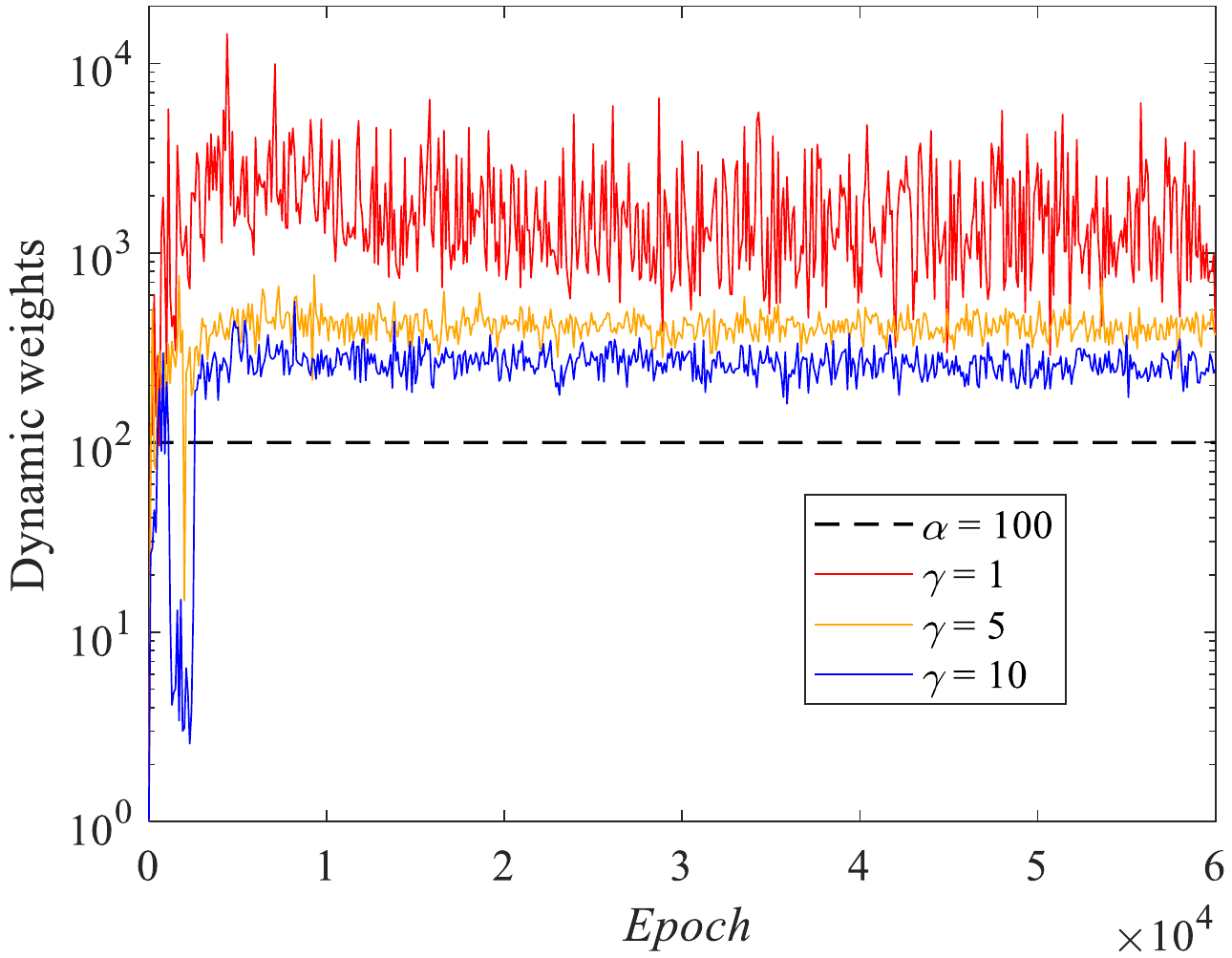}}
	\caption{Influence of weights: dynamical weights with different normalization factors versus training epochs. All the dynamic weights here are given by Eq. (\ref{e:dyW_dns}).}
	\label{fig:hyperWeights}
\end{figure}

\begin{figure}
	\centerline{\includegraphics [width=12cm] {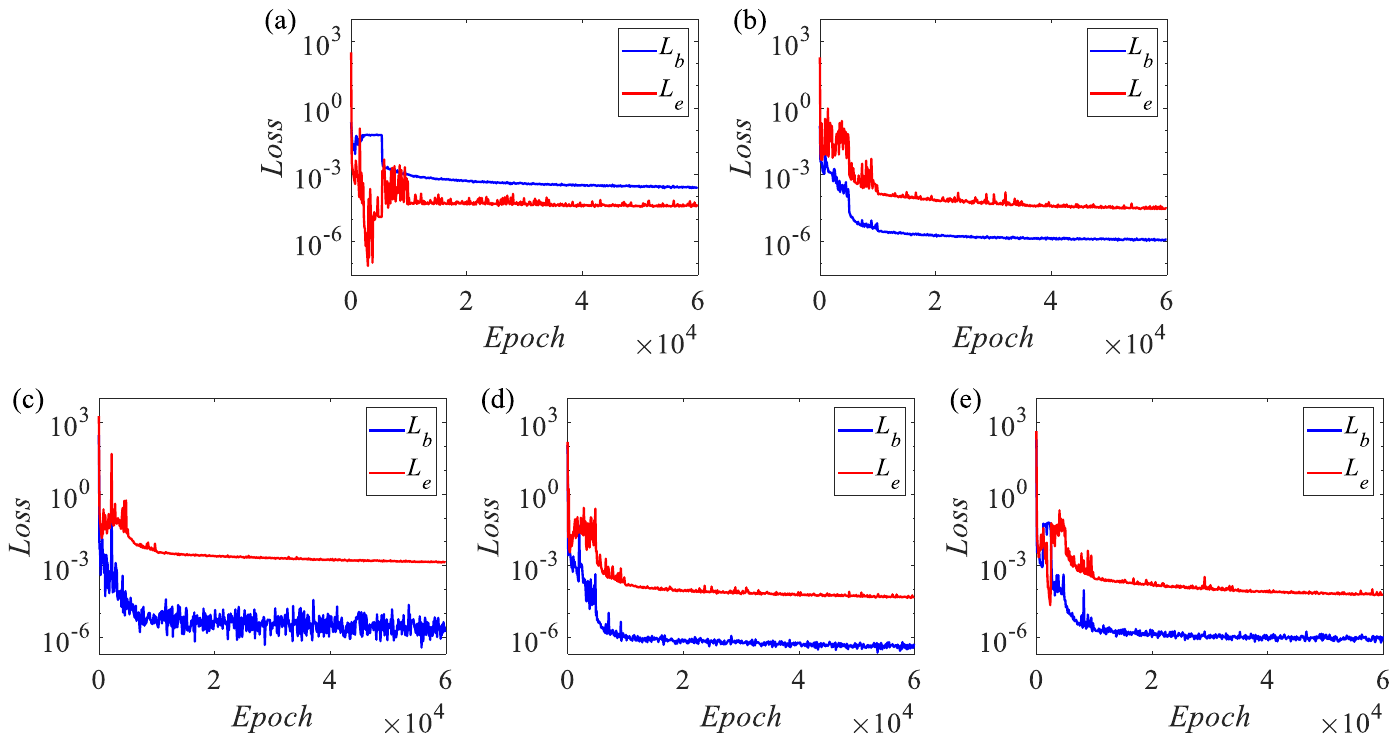}}
	\caption{Influence of weights: convergence of the VP-NSFnet simulation: (a) VP-NSFnet, fixed weight $\alpha = 1$; (b) VP-NSFnet, fixed weight $\alpha = 100$; (c) VP-NSFnet, dynamic weight with normalization factor $\gamma = 1$; (d) VP-NSFnet, dynamic weight with normalization factor $\gamma = 5$; (e) VP-NSFnet, dynamic weight with normalization factor $\gamma = 10$. All the dynamic weights here are given by Eq. (\ref{e:dyW_dns}).}
	\label{fig:hyperLoss}
\end{figure}

\begin{figure}
	\centerline{\includegraphics [width=12cm] {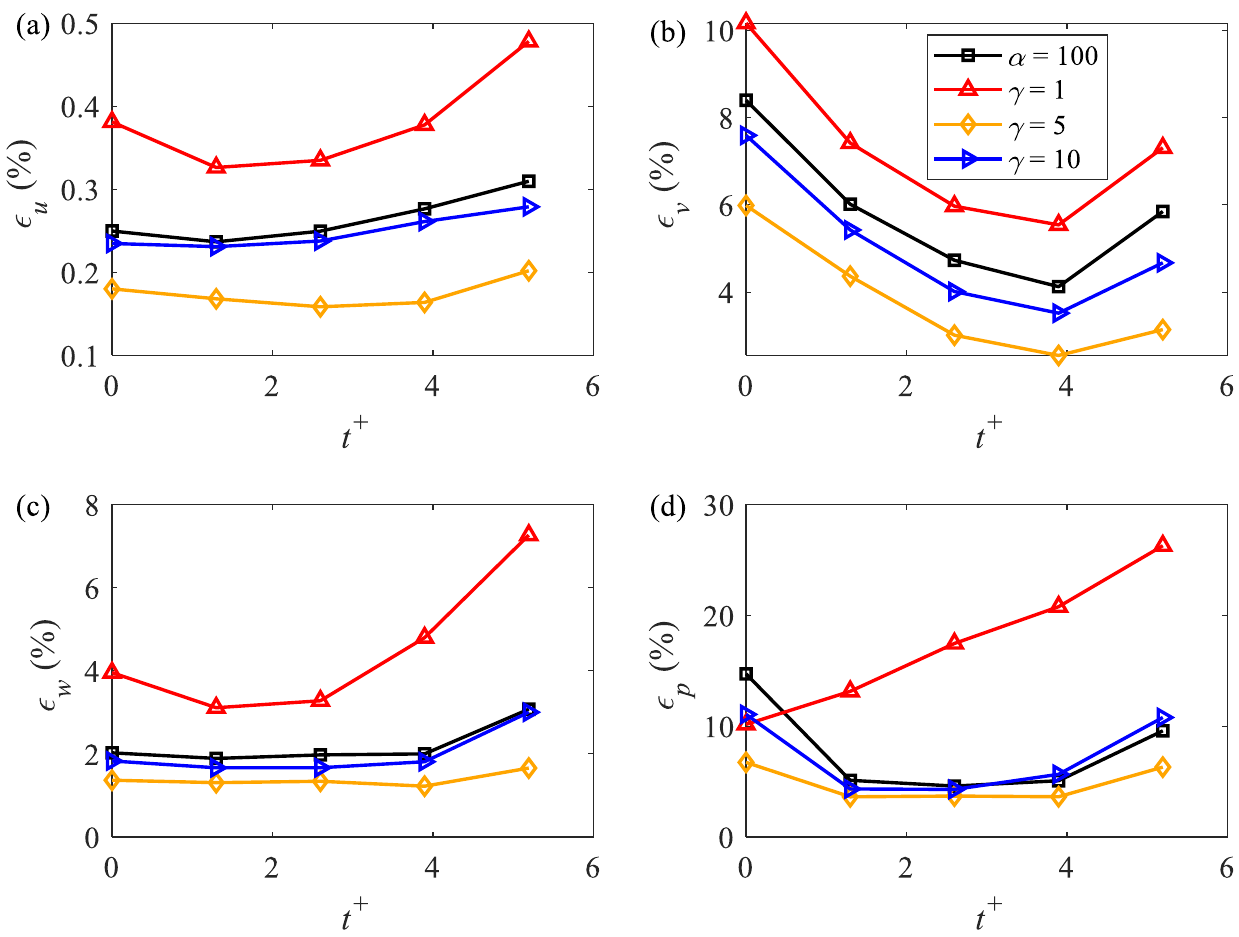}}
	\caption{Influence of weights: accuracy of VP-NSFnet for various weights: (a) to (d) relative $L_2$ errors of $u$, $v$, $w$ and $p$.}
	\label{fig:hyperErr}
\end{figure}

In the above numerical experiments for turbulent channel flow, all the weights are manually tuned to obtain satisfying results. In this section, we investigate the influence of weights, especially the dynamic weights, to the VP-NSFnet simulation accuracy. 
The formulation of Eq. (\ref{e:dyW_1}) is applied to the loss function of VP-NSFnet. However, different from Eq. (\ref{e:dyW_1}), a normalization factor $\gamma$ is taken into account in this section. Hence, the dynamic weights for the turbulence simulation can be expressed as: \begin{equation}\label{e:dyW_dns}
    \hat{\alpha}^{(k+1)} = \frac{\max_{\theta}\{|\nabla_{\theta}L_e |\}}{ \gamma\overline{|\nabla_{\theta} {\alpha}^{(k)} L_b|} }.  
\end{equation}
We consider a subdomain of [12.53, 12.59] $ \times $ [-1, -0.9762] $ \times $ [4.69, 4.75] (about 60 $ \times $ 24 $ \times $ 60 in wall-units) as the VP-NSFnet simulation domain; the non-dimensional time domain is set as [0, 0.104] (17 time steps, 5.19 in wall-units). For this small domain, the initial velocity values are not used in the loss function, i.e., $\beta = 0$, but instead we can learn them. There are 2,000 points inside the domain and 1,100 points on the boundary sampled at each time step to determine the loss function. We take the total number of iterations $n_{it}$ of one training epoch as 10. There are 5 hidden layers in the VP-NSFnet, with 200 neurons per layer. In all the examples, the initial learning rate for \emph{Adam} decays from $ 10^{-3} $ (5,000 training epochs) to $ 10^{-4} $ (5,000 training epochs), $ 10^{-5} $ (25,000 training epochs) and $ 5 \times 10^{-6} $ (25,000 training epochs). 
We set five different strategies to take different weights for the boundary data in this experiment. In the first two strategies, we use fixed weights of $\alpha = 1$ and $\alpha = 100$. Then, we use the dynamic weights given by Eq. (\ref{e:dyW_dns}) with different normalization factors, i.e., $\gamma = 1$, $\gamma = 5$ and $\gamma = 10$. The comparisons of instantaneous flow fields ($ z-y $ plane) between reference DNS and VP-NSFnet simulations at $ t^+ $ = 5.19 and $x^+ = -12.27$ with various weights are shown in Fig. \ref{fig:hyperContour}.
The evolution of dynamic weights in the training process is shown in Fig. \ref{fig:hyperWeights}.
The convergence of the VP-NSFnet simulation with various hyperparameters is shown in Fig. \ref{fig:hyperLoss}.
The accuracy of the VP-NSFnet simulation with various hyperparameters is shown in Fig. \ref{fig:hyperErr}. It is noted from Fig. \ref{fig:hyperContour} that there exists large discrepancy between the reference solution and the VP-NSFnet solution with fixed weight $\alpha=1$, thereby its accuracy is of low level. Therefore, the VP-NSFnet simulation accuracy with fixed weight $\alpha=1$ is not shown in Fig. \ref{fig:hyperErr}. 
Overall, the best result for this turbulence simulation is obtained when applying dynamic weights with $\gamma=5$. A dynamic variation of the weights can improve the performance of VP-NSFnet.

\section{Summary}\label{sec:Conclusion}
In this study, we explored the effectiveness of PINNs to directly simulate incompressible flows, ranging from laminar flows to turbulent channel flow. We have formulated NSFnets based on two different forms of the governing Navier-Stokes equations: the velocity-pressure (VP) form and the velocity-vorticity (VV) form. The spatial and temporal coordinates are the inputs of the PINNs, and the instantaneous velocity and pressure fields are the outputs for the VP-NSFnet; similarly, the instantaneous velocity and vorticity fields  are the outputs for the VV-NSFnet. We used automatic differentiation to represent all the  differential operators in the Navier-Stokes equations; then the equations can be formulated by the neural networks. We regard the initial and boundary conditions as supervised data-driven parts, and the residual of the Navier-Stokes equations as the unsupervised physics-informed part in the loss function of PINNs. We note that no data was provided for the pressure as boundary or initial conditions for VP-NSFnet, which was a hidden state and was obtained indirectly via the incompressibility constraint without splitting the equations. Convergence of NSFnets was monitored using the total loss function as well as the individual loss functions. We simulated several laminar flows, including two-dimensional steady Kovasznay flow, two-dimensional cylinder wake and three-dimensional Beltrami flow using the two forms NSFnets. We also carried out a study on the influence of weights in the various contributions to the loss function.  We found that for laminar cases, the VV-NSFnet achieves better accuracy than the VP-NSFnet, and that a dynamic variation of the  weights can improve the performance of both NSFnets.

In addition, we explored the possibility of simulating turbulent channel flow at $ \rm Re_\tau \sim $ 1,000 using NSFnets. We performed NSFnet simulations by considering different subdomains with different sizes at various locations in the channel and for different time intervals. Established DNS databases provided the proper initial and boundary conditions for the NSFnet simulations. We obtained good agreement between the DNS results and the VP-NSFnet simulation results upon convergence of the loss function. The long time period simulation suggests that NSFnets can sustain turbulence with errors bounded at reasonable levels. We also investigated the use of dynamic versus fixed weights and demonstrated how dynamic weights could further enhance the accuracy of VP-NSFnet. Unlike the VP-form, attempts to train the VV-NSFnet failed to provide satisfactory  convergence of the loss function for the governing equations. For $\alpha=50,000$ we obtained reasonable accuracy, with the loss function for boundary conditions converging to small values but the residual for the governing equations remained very large. This may be related to the fact that the data in this case  are obtained from DNS data bases with a VP formulation, and hence the boundary conditions derived may not be so accurate but rather inconsistent with the VV form of the governing equations. We plan to revisit this issue in future work.

The current study for modeling turbulence using NSFnets is the first attempt to evaluate the PINN performance, and while the first results are encouraging, the broader question is if PINNs can provide sufficient accuracy to sustain turbulence with non-stochastic boundary conditions but rather the simple zero Dirichlet and periodic boundary conditions employed for the entire domain in spectral type simulations of turbulence. To address this question expeditiously, the efficiency of PINNs has to be improved significantly, including the development of a multi-node GPU code that can accelerate significantly the training process. Such speed up can be further enhanced by using adaptive activation functions as demonstrated in \cite{jagtap2020adaptive,jagtap2019locally}.
Moreover, a further study is needed to derive the proper normalization procedure for the three different velocity components so that we can obtain uniform accuracy, as in the current study the streamwise component is inferred with an order of magnitude higher accuracy compared to the crossflow velocity components when the streamwise velocity is of an order higher than the crossflow velocity.

\section*{Acknowledgements}
X. Jin and H. Li would like to acknowledge funding from the National Natural Science Foundation of China (NSFC) (Grant No. U1711265). S. Cai and G.E. Karniadakis would like to acknowledge funding from DARPA-AIRA grant HR00111990025 and the DOE grant DE-AC05-76RL01830.

\bibliography{NSFnetsRef}

\begin{thebibliography}{33}
\expandafter\ifx\csname natexlab\endcsname\relax\def\natexlab#1{#1}\fi
\providecommand{\url}[1]{\texttt{#1}}
\providecommand{\href}[2]{#2}
\providecommand{\path}[1]{#1}
\providecommand{\DOIprefix}{doi:}
\providecommand{\ArXivprefix}{arXiv:}
\providecommand{\URLprefix}{URL: }
\providecommand{\Pubmedprefix}{pmid:}
\providecommand{\doi}[1]{\href{http://dx.doi.org/#1}{\path{#1}}}
\providecommand{\Pubmed}[1]{\href{pmid:#1}{\path{#1}}}
\providecommand{\bibinfo}[2]{#2}
\ifx\xfnm\relax \def\xfnm[#1]{\unskip,\space#1}\fi
\bibitem[{Ling et~al.(2016)Ling, Kurzawski, and Templeton}]{ling2016reynolds}
\bibinfo{author}{J.~Ling}, \bibinfo{author}{A.~Kurzawski},
  \bibinfo{author}{J.~Templeton},
\newblock \bibinfo{title}{Reynolds averaged turbulence modelling using deep
  neural networks with embedded invariance},
\newblock \bibinfo{journal}{Journal of Fluid Mechanics} \bibinfo{volume}{807}
  (\bibinfo{year}{2016}) \bibinfo{pages}{155--166}.
\bibitem[{Wang et~al.(2017)Wang, Wu, and Xiao}]{wang2017physics}
\bibinfo{author}{J.-X. Wang}, \bibinfo{author}{J.-L. Wu},
  \bibinfo{author}{H.~Xiao},
\newblock \bibinfo{title}{{Physics-informed machine learning approach for
  reconstructing Reynolds stress modeling discrepancies based on DNS data}},
\newblock \bibinfo{journal}{Physical Review Fluids} \bibinfo{volume}{2}
  (\bibinfo{year}{2017}) \bibinfo{pages}{034603}.
\bibitem[{Jiang et~al.(2020)Jiang, Mi, Laima, and Li}]{jiang2020novel}
\bibinfo{author}{C.~Jiang}, \bibinfo{author}{J.~Mi},
  \bibinfo{author}{S.~Laima}, \bibinfo{author}{H.~Li},
\newblock \bibinfo{title}{A novel algebraic stress model with
  machine-learning-assisted parameterization},
\newblock \bibinfo{journal}{Energies} \bibinfo{volume}{13}
  (\bibinfo{year}{2020}) \bibinfo{pages}{258}.
\bibitem[{Zhou et~al.(2019)Zhou, He, Wang, and Jin}]{zhou2019subgrid}
\bibinfo{author}{Z.~Zhou}, \bibinfo{author}{G.~He}, \bibinfo{author}{S.~Wang},
  \bibinfo{author}{G.~Jin},
\newblock \bibinfo{title}{Subgrid-scale model for large-eddy simulation of
  isotropic turbulent flows using an artificial neural network},
\newblock \bibinfo{journal}{Computers \& Fluids} \bibinfo{volume}{195}
  (\bibinfo{year}{2019}) \bibinfo{pages}{104319}.
\bibitem[{Jin et~al.(2018)Jin, Cheng, Chen, and Li}]{jin2018prediction}
\bibinfo{author}{X.~Jin}, \bibinfo{author}{P.~Cheng}, \bibinfo{author}{W.-L.
  Chen}, \bibinfo{author}{H.~Li},
\newblock \bibinfo{title}{{Prediction model of velocity field around circular
  cylinder over various Reynolds numbers by fusion convolutional neural
  networks based on pressure on the cylinder}},
\newblock \bibinfo{journal}{Physics of Fluids} \bibinfo{volume}{30}
  (\bibinfo{year}{2018}) \bibinfo{pages}{047105}.
\bibitem[{Wu et~al.(2020)Wu, Sun, Chang, Zhang, Arcucci, Guo, and
  Pain}]{wu2020data}
\bibinfo{author}{P.~Wu}, \bibinfo{author}{J.~Sun}, \bibinfo{author}{X.~Chang},
  \bibinfo{author}{W.~Zhang}, \bibinfo{author}{R.~Arcucci},
  \bibinfo{author}{Y.~Guo}, \bibinfo{author}{C.~C. Pain},
\newblock \bibinfo{title}{Data-driven reduced order model with temporal
  convolutional neural network},
\newblock \bibinfo{journal}{Computer Methods in Applied Mechanics and
  Engineering} \bibinfo{volume}{360} (\bibinfo{year}{2020})
  \bibinfo{pages}{112766}.
\bibitem[{Jin et~al.(2020)Jin, Laima, Chen, and Li}]{jin2020time}
\bibinfo{author}{X.~Jin}, \bibinfo{author}{S.~Laima}, \bibinfo{author}{W.-L.
  Chen}, \bibinfo{author}{H.~Li},
\newblock \bibinfo{title}{Time-resolved reconstruction of flow field around a
  circular cylinder by recurrent neural networks based on non-time-resolved
  particle image velocimetry measurements},
\newblock \bibinfo{journal}{Experiments in Fluids}  (\bibinfo{year}{2020})
  \bibinfo{pages}{accepted}.
\bibitem[{Hosseini et~al.(2015)Hosseini, Martinuzzi, and
  Noack}]{hosseini2015sensor}
\bibinfo{author}{Z.~Hosseini}, \bibinfo{author}{R.~J. Martinuzzi},
  \bibinfo{author}{B.~R. Noack},
\newblock \bibinfo{title}{Sensor-based estimation of the velocity in the wake
  of a low-aspect-ratio pyramid},
\newblock \bibinfo{journal}{Experiments in Fluids} \bibinfo{volume}{56}
  (\bibinfo{year}{2015}) \bibinfo{pages}{13}.
\bibitem[{Discetti et~al.(2018)Discetti, Raiola, and
  Ianiro}]{discetti2018estimation}
\bibinfo{author}{S.~Discetti}, \bibinfo{author}{M.~Raiola},
  \bibinfo{author}{A.~Ianiro},
\newblock \bibinfo{title}{Estimation of time-resolved turbulent fields through
  correlation of non-time-resolved field measurements and time-resolved point
  measurements},
\newblock \bibinfo{journal}{Experimental Thermal and Fluid Science}
  \bibinfo{volume}{93} (\bibinfo{year}{2018}) \bibinfo{pages}{119--130}.
\bibitem[{Cai et~al.(2019)Cai, Zhou, Xu, and Gao}]{cai2019dense}
\bibinfo{author}{S.~Cai}, \bibinfo{author}{S.~Zhou}, \bibinfo{author}{C.~Xu},
  \bibinfo{author}{Q.~Gao},
\newblock \bibinfo{title}{Dense motion estimation of particle images via a
  convolutional neural network},
\newblock \bibinfo{journal}{Experiments in Fluids} \bibinfo{volume}{60}
  (\bibinfo{year}{2019}) \bibinfo{pages}{73}.
\bibitem[{Duraisamy et~al.(2019)Duraisamy, Iaccarino, and
  Xiao}]{duraisamy2019turbulence}
\bibinfo{author}{K.~Duraisamy}, \bibinfo{author}{G.~Iaccarino},
  \bibinfo{author}{H.~Xiao},
\newblock \bibinfo{title}{Turbulence modeling in the age of data},
\newblock \bibinfo{journal}{Annual Review of Fluid Mechanics}
  \bibinfo{volume}{51} (\bibinfo{year}{2019}) \bibinfo{pages}{357--377}.
\bibitem[{Brunton et~al.(2020)Brunton, Noack, and
  Koumoutsakos}]{brunton2020machine}
\bibinfo{author}{S.~L. Brunton}, \bibinfo{author}{B.~R. Noack},
  \bibinfo{author}{P.~Koumoutsakos},
\newblock \bibinfo{title}{Machine learning for fluid mechanics},
\newblock \bibinfo{journal}{Annual Review of Fluid Mechanics}
  \bibinfo{volume}{52} (\bibinfo{year}{2020}) \bibinfo{pages}{477--508}.
\bibitem[{Raissi et~al.(2017{\natexlab{a}})Raissi, Perdikaris, and
  Karniadakis}]{raissi2017physicsA}
\bibinfo{author}{M.~Raissi}, \bibinfo{author}{P.~Perdikaris},
  \bibinfo{author}{G.~E. Karniadakis},
\newblock \bibinfo{title}{Physics informed deep learning (part i): Data-driven
  solutions of nonlinear partial differential equations},
\newblock \bibinfo{journal}{arXiv preprint arXiv:1711.10561}
  (\bibinfo{year}{2017}{\natexlab{a}}).
\bibitem[{Raissi et~al.(2017{\natexlab{b}})Raissi, Perdikaris, and
  Karniadakis}]{raissi2017physicsB}
\bibinfo{author}{M.~Raissi}, \bibinfo{author}{P.~Perdikaris},
  \bibinfo{author}{G.~E. Karniadakis},
\newblock \bibinfo{title}{Physics informed deep learning (part ii): Data-driven
  discovery of nonlinear partial differential equations. arxiv},
\newblock \bibinfo{journal}{arXiv preprint arXiv:1711.10561}
  (\bibinfo{year}{2017}{\natexlab{b}}).
\bibitem[{Raissi et~al.(2019{\natexlab{a}})Raissi, Perdikaris, and
  Karniadakis}]{raissi2019physics}
\bibinfo{author}{M.~Raissi}, \bibinfo{author}{P.~Perdikaris},
  \bibinfo{author}{G.~E. Karniadakis},
\newblock \bibinfo{title}{Physics-informed neural networks: A deep learning
  framework for solving forward and inverse problems involving nonlinear
  partial differential equations},
\newblock \bibinfo{journal}{Journal of Computational Physics}
  \bibinfo{volume}{378} (\bibinfo{year}{2019}{\natexlab{a}})
  \bibinfo{pages}{686--707}.
\bibitem[{Raissi et~al.(2019{\natexlab{b}})Raissi, Wang, Triantafyllou, and
  Karniadakis}]{raissi2019deep}
\bibinfo{author}{M.~Raissi}, \bibinfo{author}{Z.~Wang}, \bibinfo{author}{M.~S.
  Triantafyllou}, \bibinfo{author}{G.~E. Karniadakis},
\newblock \bibinfo{title}{Deep learning of vortex-induced vibrations},
\newblock \bibinfo{journal}{Journal of Fluid Mechanics} \bibinfo{volume}{861}
  (\bibinfo{year}{2019}{\natexlab{b}}) \bibinfo{pages}{119--137}.
\bibitem[{Raissi et~al.(2020)Raissi, Yazdani, and
  Karniadakis}]{raissi2020hidden}
\bibinfo{author}{M.~Raissi}, \bibinfo{author}{A.~Yazdani},
  \bibinfo{author}{G.~E. Karniadakis},
\newblock \bibinfo{title}{Hidden fluid mechanics: Learning velocity and
  pressure fields from flow visualizations},
\newblock \bibinfo{journal}{Science} \bibinfo{volume}{367}
  (\bibinfo{year}{2020}) \bibinfo{pages}{1026--1030}.
\bibitem[{Kim et~al.(1987)Kim, Moin, and Moser}]{kim1987turbulence}
\bibinfo{author}{J.~Kim}, \bibinfo{author}{P.~Moin},
  \bibinfo{author}{R.~Moser},
\newblock \bibinfo{title}{{Turbulence statistics in fully developed channel
  flow at low Reynolds number}},
\newblock \bibinfo{journal}{Journal of Fluid Mechanics} \bibinfo{volume}{177}
  (\bibinfo{year}{1987}) \bibinfo{pages}{133--166}.
\bibitem[{Karniadakis and Sherwin(2013)}]{karniadakis2013spectral}
\bibinfo{author}{G.~E. Karniadakis}, \bibinfo{author}{S.~Sherwin},
  \bibinfo{title}{Spectral/hp Element Methods for Computational Fluid
  Dynamics}, \bibinfo{publisher}{Oxford University Press},
  \bibinfo{year}{2013}.
\bibitem[{Perlman et~al.(2007)Perlman, Burns, Li, and
  Meneveau}]{perlman2007data}
\bibinfo{author}{E.~Perlman}, \bibinfo{author}{R.~Burns},
  \bibinfo{author}{Y.~Li}, \bibinfo{author}{C.~Meneveau},
\newblock \bibinfo{title}{Data exploration of turbulence simulations using a
  database cluster},
\newblock in: \bibinfo{booktitle}{Proceedings of the 2007 ACM/IEEE conference
  on Supercomputing}, \bibinfo{organization}{ACM}, \bibinfo{year}{2007},
  p.~\bibinfo{pages}{23}.
\bibitem[{Li et~al.(2008)Li, Perlman, Wan, Yang, Meneveau, Burns, Chen, Szalay,
  and Eyink}]{li2008public}
\bibinfo{author}{Y.~Li}, \bibinfo{author}{E.~Perlman},
  \bibinfo{author}{M.~Wan}, \bibinfo{author}{Y.~Yang},
  \bibinfo{author}{C.~Meneveau}, \bibinfo{author}{R.~Burns},
  \bibinfo{author}{S.~Chen}, \bibinfo{author}{A.~Szalay},
  \bibinfo{author}{G.~Eyink},
\newblock \bibinfo{title}{{A public turbulence database cluster and
  applications to study Lagrangian evolution of velocity increments in
  turbulence}},
\newblock \bibinfo{journal}{Journal of Turbulence} \bibinfo{volume}{9}
  (\bibinfo{year}{2008}) \bibinfo{pages}{N31}.
\bibitem[{Graham et~al.(2016)Graham, Kanov, Yang, Lee, Malaya, Lalescu, Burns,
  Eyink, Szalay, Moser et~al.}]{graham2016web}
\bibinfo{author}{J.~Graham}, \bibinfo{author}{K.~Kanov},
  \bibinfo{author}{X.~Yang}, \bibinfo{author}{M.~Lee},
  \bibinfo{author}{N.~Malaya}, \bibinfo{author}{C.~Lalescu},
  \bibinfo{author}{R.~Burns}, \bibinfo{author}{G.~Eyink},
  \bibinfo{author}{A.~Szalay}, \bibinfo{author}{R.~Moser}, et~al.,
\newblock \bibinfo{title}{A web services accessible database of turbulent
  channel flow and its use for testing a new integral wall model for les},
\newblock \bibinfo{journal}{Journal of Turbulence} \bibinfo{volume}{17}
  (\bibinfo{year}{2016}) \bibinfo{pages}{181--215}.
\bibitem[{Baydin et~al.(2018)Baydin, Pearlmutter, Radul, and
  Siskind}]{baydin2018automatic}
\bibinfo{author}{A.~G. Baydin}, \bibinfo{author}{B.~A. Pearlmutter},
  \bibinfo{author}{A.~A. Radul}, \bibinfo{author}{J.~M. Siskind},
\newblock \bibinfo{title}{Automatic differentiation in machine learning: a
  survey},
\newblock \bibinfo{journal}{Journal of Machine Learning Research}
  \bibinfo{volume}{18} (\bibinfo{year}{2018}).
\bibitem[{Karniadakis et~al.(1991)Karniadakis, Israeli, and
  Orszag}]{karniadakis1991high}
\bibinfo{author}{G.~E. Karniadakis}, \bibinfo{author}{M.~Israeli},
  \bibinfo{author}{S.~A. Orszag},
\newblock \bibinfo{title}{High-order splitting methods for the incompressible
  navier-stokes equations},
\newblock \bibinfo{journal}{Journal of Computational Physics}
  \bibinfo{volume}{97} (\bibinfo{year}{1991}) \bibinfo{pages}{414--443}.
\bibitem[{Wang et~al.(2020)Wang, Teng, and Perdikaris}]{wang2020understanding}
\bibinfo{author}{S.~Wang}, \bibinfo{author}{Y.~Teng},
  \bibinfo{author}{P.~Perdikaris},
\newblock \bibinfo{title}{Understanding and mitigating gradient pathologies in
  physics-informed neural networks},
\newblock \bibinfo{journal}{arXiv preprint arXiv:2001.04536}
  (\bibinfo{year}{2020}).
\bibitem[{Kingma and Ba(2014)}]{kingma2014adam}
\bibinfo{author}{D.~P. Kingma}, \bibinfo{author}{J.~Ba},
\newblock \bibinfo{title}{Adam: A method for stochastic optimization},
\newblock \bibinfo{journal}{arXiv preprint arXiv:1412.6980}
  (\bibinfo{year}{2014}).
\bibitem[{Glorot and Bengio(2010)}]{glorot2010understanding}
\bibinfo{author}{X.~Glorot}, \bibinfo{author}{Y.~Bengio},
\newblock \bibinfo{title}{Understanding the difficulty of training deep
  feedforward neural networks},
\newblock in: \bibinfo{booktitle}{Proceedings of the thirteenth international
  conference on artificial intelligence and statistics}, \bibinfo{year}{2010},
  pp. \bibinfo{pages}{249--256}.
\bibitem[{Trujillo and Karniadakis(1999)}]{trujillo1999penalty}
\bibinfo{author}{J.~Trujillo}, \bibinfo{author}{G.~E. Karniadakis},
\newblock \bibinfo{title}{A penalty method for the vorticity-velocity
  formulation},
\newblock \bibinfo{journal}{Journal of Computational Physics}
  \bibinfo{volume}{149} (\bibinfo{year}{1999}) \bibinfo{pages}{32--58}.
\bibitem[{Meitz and Fasel(2000)}]{meitz2000compact}
\bibinfo{author}{H.~L. Meitz}, \bibinfo{author}{H.~F. Fasel},
\newblock \bibinfo{title}{{A compact-difference scheme for the Navier--Stokes
  equations in vorticity-velocity formulation}},
\newblock \bibinfo{journal}{Journal of Computational Physics}
  \bibinfo{volume}{157} (\bibinfo{year}{2000}) \bibinfo{pages}{371--403}.
\bibitem[{Jagtap et~al.(2020)Jagtap, Kawaguchi, and
  Karniadakis}]{jagtap2020adaptive}
\bibinfo{author}{A.~D. Jagtap}, \bibinfo{author}{K.~Kawaguchi},
  \bibinfo{author}{G.~E. Karniadakis},
\newblock \bibinfo{title}{Adaptive activation functions accelerate convergence
  in deep and physics-informed neural networks},
\newblock \bibinfo{journal}{Journal of Computational Physics}
  \bibinfo{volume}{404} (\bibinfo{year}{2020}) \bibinfo{pages}{109136}.
\bibitem[{Jagtap et~al.(2019)Jagtap, Kawaguchi, and
  Karniadakis}]{jagtap2019locally}
\bibinfo{author}{A.~D. Jagtap}, \bibinfo{author}{K.~Kawaguchi},
  \bibinfo{author}{G.~E. Karniadakis},
\newblock \bibinfo{title}{Locally adaptive activation functions with slope
  recovery term for deep and physics-informed neural networks},
\newblock \bibinfo{journal}{arXiv preprint arXiv:1909.12228}
  (\bibinfo{year}{2019}).
\bibitem[{Ethier and Steinman(1994)}]{ethier1994exact}
\bibinfo{author}{C.~R. Ethier}, \bibinfo{author}{D.~Steinman},
\newblock \bibinfo{title}{Exact fully 3{D Navier-Stokes} solutions for
  benchmarking},
\newblock \bibinfo{journal}{International Journal for Numerical Methods in
  Fluids} \bibinfo{volume}{19} (\bibinfo{year}{1994})
  \bibinfo{pages}{369--375}.
\bibitem[{Pope(2000)}]{pope2000turbulent}
\bibinfo{author}{S.~B. Pope},
\newblock \bibinfo{title}{{Turbulent Flows}},
\newblock \bibinfo{journal}{Cambridge University Press, Cambridge.}
  (\bibinfo{year}{2000}) \bibinfo{pages}{276}.

\end{thebibliography}

\end{document}